\documentclass[aps, prd,twocolumn,superscriptaddress,nofootinbib,longbibliography,preprintnumbers]{revtex4-1}  

\usepackage[a4paper,top=20mm,bottom=20mm, width=175mm]{geometry}

\usepackage[T1]{fontenc}

\usepackage{graphicx}  
\graphicspath{{.}}
\usepackage{dcolumn}   
\usepackage{bm}        
\usepackage{amsmath}	
\usepackage{amssymb}   
\usepackage{comment} 
\usepackage{xcolor}
\usepackage{hyperref}
\usepackage[pagewise]{lineno}

\def\metacal{\textsc{Metacalibration}}
\def\im3shape{\textsc{im3shape}}
\def\healpix{\textsc{HEALPix}}
\def\flask{\textsc{FLASK}}

\newcommand\redd[1]{\textcolor{black}{#1}}

\begin{document}



\title[]{Dark Energy Survey Year 1 Results: Galaxy-Galaxy Lensing}
                             
\author{J.~Prat}
\email[Corresponding author: ]{jprat@ifae.es}
\affiliation{Institut de F\'{\i}sica d'Altes Energies (IFAE), The Barcelona Institute of Science and Technology, Campus UAB, 08193 Bellaterra (Barcelona) Spain}
\author{C.~S{\'a}nchez}
\email[Corresponding author: ]{csanchez@ifae.es}
\affiliation{Institut de F\'{\i}sica d'Altes Energies (IFAE), The Barcelona Institute of Science and Technology, Campus UAB, 08193 Bellaterra (Barcelona) Spain}
\author{Y.~Fang}
\affiliation{Department of Physics and Astronomy, University of Pennsylvania, Philadelphia, PA 19104, USA}
\author{D.~Gruen}
\affiliation{Kavli Institute for Particle Astrophysics \& Cosmology, P. O. Box 2450, Stanford University, Stanford, CA 94305, USA}
\affiliation{SLAC National Accelerator Laboratory, Menlo Park, CA 94025, USA}
\affiliation{Einstein Fellow}
\author{J.~Elvin-Poole}
\affiliation{Jodrell Bank Center for Astrophysics, School of Physics and Astronomy, University of Manchester, Oxford Road, Manchester, M13 9PL, UK}
\author{N.~Kokron}
\affiliation{Departamento de F\'isica Matem\'atica, Instituto de F\'isica, Universidade de S\~ao Paulo, CP 66318, S\~ao Paulo, SP, 05314-970, Brazil}
\affiliation{Laborat\'orio Interinstitucional de e-Astronomia - LIneA, Rua Gal. Jos\'e Cristino 77, Rio de Janeiro, RJ - 20921-400, Brazil}
\author{L.~F.~Secco}
\affiliation{Department of Physics and Astronomy, University of Pennsylvania, Philadelphia, PA 19104, USA}
\author{B.~Jain}
\affiliation{Department of Physics and Astronomy, University of Pennsylvania, Philadelphia, PA 19104, USA}
\author{R.~Miquel}
\affiliation{Instituci\'o Catalana de Recerca i Estudis Avan\c{c}ats, E-08010 Barcelona, Spain}
\affiliation{Institut de F\'{\i}sica d'Altes Energies (IFAE), The Barcelona Institute of Science and Technology, Campus UAB, 08193 Bellaterra (Barcelona) Spain}
\author{N.~MacCrann}
\affiliation{Center for Cosmology and Astro-Particle Physics, The Ohio State University, Columbus, OH 43210, USA}
\affiliation{Department of Physics, The Ohio State University, Columbus, OH 43210, USA}
% ...
\collaboration{DES Collaboration}
\date{\today}

\begin{abstract}
We present galaxy-galaxy lensing measurements from 1321 sq.~deg.~of the Dark Energy Survey (DES) Year 1 (Y1) data. The lens sample consists of a selection of 660,000 red galaxies with high-precision photometric redshifts, known as redMaGiC, split into five tomographic bins in the redshift range $0.15 < z < 0.9$. We use two different source samples, obtained from the \textsc{Metacalibration} (26 million galaxies) and \textsc{im3shape} (18 million galaxies) shear estimation codes, which are split into four photometric redshift bins in the range $0.2 < z < 1.3$. We perform extensive testing of potential systematic effects that can bias the galaxy-galaxy lensing signal, including those from shear estimation, photometric redshifts, and observational properties. Covariances are obtained from jackknife subsamples of the data and validated with a suite of log-normal simulations. We use the shear-ratio geometric test to obtain independent constraints on the mean of the source redshift distributions, providing validation of those obtained from other photo-$z$ studies with the same data. We find consistency between the galaxy bias estimates obtained from our galaxy-galaxy lensing measurements and from galaxy clustering, therefore showing the galaxy-matter cross-correlation coefficient $r$ to be consistent with one, measured over the scales used for the cosmological analysis. The results in this work present one of the three two-point correlation functions, along with galaxy clustering and cosmic shear, used in the DES cosmological analysis of Y1 data, and hence the methodology and the systematics tests presented here provide a critical input for that study as well as for future cosmological analyses in DES and other photometric galaxy surveys. 
\end{abstract}

\preprint{DES-2016-0210}
\preprint{FERMILAB-PUB-17-277-PPD}
\maketitle



\section{Introduction}

Weak gravitational lensing refers to the small distortions in the images of distant galaxies by intervening mass along the line of sight. Galaxy-galaxy lensing refers to the cross-correlation between foreground (lens) galaxy positions and the lensing shear of background (source) galaxies at higher redshifts \citep{Tyson1984,Brainerd1996,DellAntonio1996}. The component of the shear that is tangential to the perpendicular line connecting the lens and source galaxies is a measure of the projected, excess mass distribution around the lens galaxies. Galaxy-galaxy lensing at small scales has been used to characterize the properties of dark matter halos hosting lens galaxies, while at large scales it measures the cross correlation between galaxy and matter densities. The measurements have many applications, ranging from constraining halo mass profiles \cite{Navarro1997} to estimating the large-scale bias of a given galaxy population to obtaining cosmological constraints \cite{Cacciato2009,Mandelbaum2013,Cacciato2013,More2015,Kwan2016,VanUitert2017}. Recent surveys such as CFHTLenS \cite{Heymans2012,Erben2013} have presented measurements on galaxy-galaxy lensing \cite{Gillis2013, Velander2014,Hudson2015}. Similarly, measurements from KiDS \cite{DeJong2013,Kuijken2015} have also studied the galaxy-mass connection using galaxy-galaxy lensing \cite{Sifon2015,Viola2015,VanUitert2016,Joudaki2017}. The galaxy-mass connection has also been studied in \cite{Sheldon2004,Mandelbaum2006} and by \cite{Leauthaud2012} at high redshift.

In this paper we present measurements and extensive tests of the tomographic galaxy-galaxy lensing signal from Year 1 data of the Dark Energy Survey (DES). DES is an ongoing wide-field multi-band imaging survey that will cover 5000 sq.~deg.~of the Southern sky over five years. Our goals are to present the measurements of galaxy-galaxy lensing with DES, carry out a series of null tests of our measurement pipeline and the data, and carry out related analyses of the lensing and photometric redshift (photo-$z$) performance that are critical for the 
Y1 cosmological analysis \cite{keypaper}. We use five redshift bins for the lens galaxies and four bins for the source galaxies. The detailed tests presented here will serve as a foundation for future work relying on galaxy-galaxy lensing measurements, such as Halo Occupation Distribution (HOD) analyses \cite{Berlind2002,COORAY2002}. The galaxy-galaxy lensing studies with the DES Science Verification (SV) data serve as precursors to this paper \cite{Clampitt2016,Prat2016,Park2016,Kwan2016}.

The lens galaxy sample used is the red-sequence Matched-filter Galaxy Catalog (redMaGiC, \cite{Rozo2015}), which is a catalog of photometrically selected luminous red galaxies (LRGs). The redMaGiC algorithm uses the redMaPPer-calibrated model for the color of red-sequence galaxies as a function of magnitude and redshift \cite{Rykoff2014, Rykoff2016}. This algorithm constructs a galaxy sample with far more reliable redshift estimates than is achievable for a typical galaxy in DES.

For the source galaxy redshifts, we rely on less well-constrained photo-$z$ estimates, calibrated in two independent ways \cite{photoz, xcorrtechnique, xcorr}. In this paper, we use the expected behavior of the galaxy-galaxy lensing signal with the distance to source galaxies (the shear-ratio test) to validate the photo-$z$ estimates and calibration. 
The scaling of the galaxy-galaxy lensing signal with source redshift for a given lens bin is mostly driven by the geometry of the lens-source configuration, with cosmology dependence being subdominant to potential biases in the redshift estimation of the galaxies involved. Therefore, such measurements provide useful constraints on the redshift distribution of source galaxies, which we then compare to findings by independent studies. 

The DES Y1 cosmological analysis \cite{keypaper} relies on the assumption that the cross-correlation coefficient between galaxies and matter is unity on the scales used for this analysis. In this work we provide validation for this assumption by showing the linear galaxy bias estimates from galaxy-galaxy lensing to be consistent with those obtained from galaxy clustering using the same galaxy sample \cite{wthetapaper}.

The plan of the paper is as follows. In Section~\ref{sec:theory}, we present the modelling. Section~\ref{sec:data} describes our data, including basic details of DES, descriptions of the lens galaxy sample, pipelines for source galaxy shape measurements, and the photometric redshift estimation of lens and source galaxies. We also describe a set of lognormal simulations used for tests of the measurement methodology. The details of the measurement and covariance estimation, together with our galaxy-galaxy lensing measurements, are presented in Section~\ref{sec:measurement}. Tests of potential systematic effects on the measurement are shown in Section~\ref{sec:tests}. Section~\ref{sec:shearratio} presents the use of tomographic galaxy-galaxy lensing to test the photo-$z$'s of source galaxies. Finally, in Section~\ref{sec:bias} we compare the galaxy bias estimates from galaxy-galaxy lensing to those obtained using the angular clustering of galaxies \cite{wthetapaper}, and we conclude in Section~\ref{sec:conclusions}.

\section{Theory}\label{sec:theory}

Galaxy-galaxy lensing is the measurement of the tangential shear of background (source) galaxies around foreground (lens) galaxies (see \cite{Bartelmann2001} for a review). The amplitude of distortion in the shapes of source galaxies is correlated with the amount of mass 
that causes passing light rays to bend. Assuming that lens galaxies trace the mass distribution following a simple linear biasing model ($\delta_g = b \:\delta_m$), the galaxy-matter power spectrum relates to the matter power spectrum by a single multiplicative bias factor. In this case, the tangential shear of background galaxies in redshift bin $j$ around foreground galaxy positions in redshift bin $i$ at an angular separation $\theta$ can be written as the following integral over the matter power spectrum $P_{\delta \delta}$:
\begin{eqnarray}\nonumber
\gamma^{ij}_t  (\theta) =  b^i \: \frac{3}{2} \, \Omega_m  \left( \frac{H_0}{c}\right)^2  \int \frac{d\ell}{2\pi} \, \ell \, J_2(\theta \ell ) \times \\
\times \int dz \left[ \frac{g^j(z)\,n_l^i(z)}{a(z)\, \chi(z) } \,  P_{\delta \delta} \left( k= \frac{\ell}{\chi(z)}, \, \chi(z) \right)\right],
\label{eq:cosmo}
\end{eqnarray}
where we are assuming $b^i(z)=b^i$ within a lens redshift bin, $J_2$ is the second order Bessel function, $l$ is the multipole moment, $k$ is the 3D wavenumber, $a$ is the scale factor, $\chi$ is the comoving distance to redshift $z$, $n_l^i(z)$ is the redshift distribution of foreground (lens) galaxies in bin $i$ and $g^j(z)$ is the lensing efficiency for background galaxies in bin $j$, computed as
\begin{equation}
g^j(z) = \int_{z}^{\infty} dz '\, n_s^j(z') \, \frac{\chi(z') - \chi(z)}{\chi(z')},
\end{equation}
where $n_s^j(z)$ is the corresponding redshift distribution of background (source) galaxies in bin $j$. The tangential shear in Eq.~(\ref{eq:cosmo}) depends on the cosmological parameters not only through the explicit dependencies but also through the matter power spectrum $P_{\delta \delta}$. Nonetheless, the dependence on the cosmological parameters is heavily degenerate with the galaxy bias of the lens galaxy population, $b^i$.    

It is also useful to express the tangential shear in terms of the excess surface mass density $\Delta \Sigma$. \redd{This estimator is typically used to study the properties of dark matter halos (see for instance \cite{Mandelbaum2006}). However, with the large scales used in this analysis, the lensing effect is caused by general matter overdensities which are traced by galaxies. In this work, we make use of this estimator because the geometrical dependence of the lensing signal becomes more evident.} The estimator reads:
\begin{equation}\label{eq:gammat_delta_sigma}
\gamma_t = \frac{\Delta \Sigma}{\Sigma_\mathrm{crit}},
\end{equation}
where the lensing strength $\Sigma_{\mathrm{crit}}^{-1}$ is a geometrical factor that depends on the angular diameter distance to the lens $D_{\rm l}$, the source $D_{\rm s}$ and the relative distance between them $D_{\rm ls}$:
\begin{equation}\label{eq:inverse_sigma_crit}
\Sigma_{\mathrm{crit}}^{-1} (z_{\rm l}, z_{\rm s}) = \frac{4\pi G}{c^2} \frac{D_{\rm ls} \, D_{\rm l}}{D_{\rm s}},
\end{equation}
with $\Sigma_{\mathrm{crit}}^{-1}(z_l,z_s)=0$ for $z_s<z_l$, and where $z_l$ and $z_s$ are the lens and source galaxy redshifts, respectively. Since the redshift distributions of our lens and source samples, $n_l(z), n_s(z)$ respectively, have a non-negligible width and even overlap, we take this into account by defining an effective $\Sigma^{-1}_{\mathrm{crit}}$ integrating over the corresponding redshift distributions. For a given lens bin $i$ and source bin $j$, this has the following form:
\begin{equation}\label{eq:eff_inverse_sigma_crit}
\Sigma_{\mathrm{crit},\mathrm{eff}}^{-1\ i,j} = \int \int dz_l dz_s \, n_l^i(z_l) \, n_s^j(z_s) \, \Sigma_{\mathrm{crit}}^{-1}(z_l, z_s).
\end{equation}

We need to assume a certain cosmology (flat $\Lambda$CDM with $\Omega_m = 0.3$) when calculating the angular diameter distances in $\Sigma^{-1}_{\mathrm{crit}}$. The results presented in this analysis depend only weakly on this choice of cosmology, as we will further discuss in the relevant sections (see Sec.~\ref{sec:shearratio}). 

\section{Data and simulations}
\label{sec:data}

\begin{figure}
 \includegraphics[width=\columnwidth]{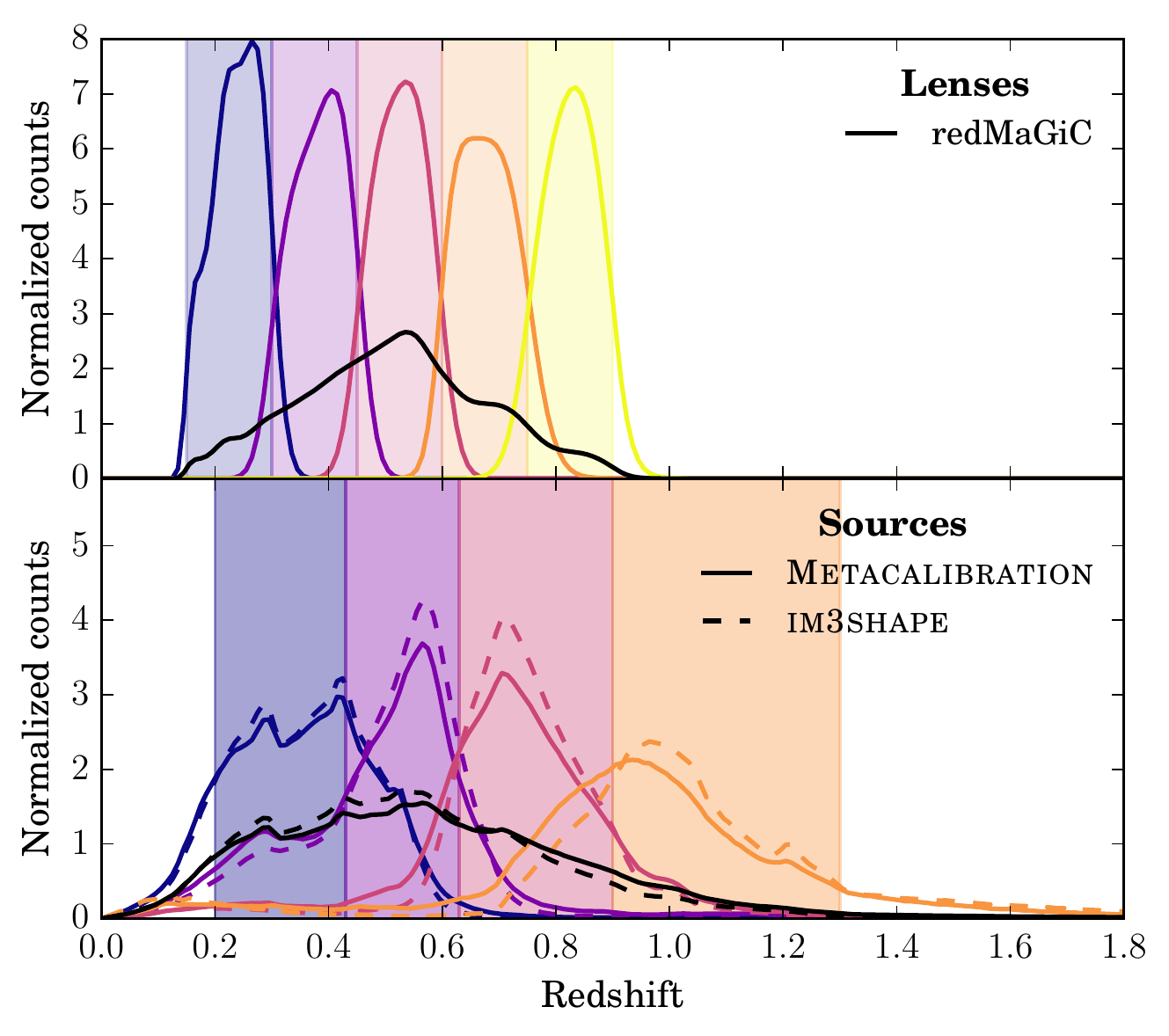}
 \caption{(\textit{Top panel}): Redshift distributions of redMaGiC lens galaxies divided in tomographic bins (\textit{colors}) and for the combination of all of them (\textit{black}). The $n(z)$'s are obtained stacking individual Gaussian distributions for each galaxy. (\textit{Bottom panel}): The same, but for our two weak lensing source samples, \textsc{Metacalibration} and \textsc{im3shape}, using the BPZ photometric redshift code.}
 \label{fig:N(z)}
\end{figure}

The Dark Energy Survey is a photometric survey that will cover about one 
quarter  
of the southern sky (5000 sq.~deg.) to a depth of $r>24$, imaging about 300 million galaxies in 5 broadband filters ($grizY$) up to redshift $z = 1.4$ \cite{Flaugher2015,DES2016}. In this work we use data from 
a large contiguous region of 1321 sq.~deg.~of DES Year 1 observations which overlaps with the South Pole Telescope footprint $-60$~deg.~< $\delta$ < $-40$~deg.~and reaches a limiting magnitude of $\approx23$ in the $r$-band (with a mean of 3 exposures out of the planned 10 for the full survey). Y1 images were taken between 31 Aug 2013 and 9 Feb 2014. 

\subsection{Lens sample: redMaGiC}
\label{subsec:redmagic}

The lens galaxy sample used in this work is a subset of the DES Y1 Gold Catalog \cite{y1gold} selected by redMaGiC \cite{Rozo2015}, which is an algorithm designed to define a sample of luminous red galaxies (LRGs) with minimal photo-$z$ uncertainties. It selects galaxies above some luminosity threshold based on how well they fit a red sequence template, calibrated using redMaPPer \cite{Rykoff2014,Rykoff2016} and a subset of galaxies with spectroscopically verified redshifts. The cutoff in the goodness of fit to the red sequence is imposed as a function of redshift and adjusted such that a constant comoving number density of galaxies is maintained.
The redMaGiC photo-$z$'s show excellent performance, with a scatter of $\sigma_z/(1+z) = 0.0166$ \cite{wthetapaper}. Furthermore, their errors are very well characterized and approximately Gaussian, enabling the redshift distribution of a sample, $n(z)$, to be obtained by stacking each galaxy's Gaussian redshift probability distribution function (see \cite{Rozo2015} for more details).

The sample used in this work is a combination of three redMaGiC galaxy samples, each of them defined to be complete down to a given luminosity threshold $L_{\rm min}$. We split the lens sample into five equally-spaced tomographic redshift bins between $z = 0.15$ and $z = 0.9$, with the three lower redshift bins using the lowest luminosity threshold 
of $L_{\rm min}=0.5L^{\star}$ (named High Density sample) and the two highest redshift bins using higher luminosity thresholds 
of $L_{\rm min}=1.0L^{\star}$  and $L_{\rm min}=1.5L^{\star}$ (named High Luminosity and Higher Luminosity samples, respectively). 
Using the stacking procedure mentioned above, redshift distributions are obtained  and shown in Fig.~\ref{fig:N(z)}. Furthermore, redMaGiC samples have been produced with two different photometric reduction techniques, \texttt{MAG\_AUTO} and Multi-object fitting photometry (\texttt{MOF}), both described in \cite{y1gold}. We follow the analysis of \cite{wthetapaper} and we use \texttt{MAG\_AUTO} photometry for the three lower redshift bins and \texttt{MOF} photometry for the rest, as it was found in \cite{wthetapaper} that this combination was optimal in minimizing systematic effects that 
introduce spurious angular galaxy clustering.

\subsection{Source samples: \textsc{Metacalibration} and \textsc{im3shape}}
\label{subsec:sources}

\textsc{Metacalibration} \cite{Huff2017,Sheldon2017} is a recently developed method to accurately measure weak lensing shear using only the available imaging data, without need for prior information about galaxy properties or calibration from simulations. The method involves distorting the image with a small known shear, and calculating the response of a shear estimator to that applied shear. This new technique
can be applied to any shear estimation code provided it fulfills certain requirements. For this work, it has been applied to the \textsc{ngmix} shear pipeline \cite{Sheldon2014}, which fits a Gaussian model simultaneously in the $riz$ bands to measure the ellipticities of the galaxies. The details of this implementation can be found in  \cite{shearcat}. We will refer to the \textsc{ngmix} shear catalog calibrated using that procedure as \metacal.

\im3shape is based on the algorithm by \cite{Zuntz2013}, modified according to \cite{Jarvis2015} and \cite{shearcat}. It performs a maximum likelihood fit using a bulge-or-disk galaxy model to estimate the ellipticity of a galaxy, i.e.~it fits de Vaucouleurs bulge and exponential disk components to galaxy images in the $r$ band, with shear biases calibrated from realistic simulations \cite{shearcat,y1-neighbours}.

Due to conservative cuts on measured galaxy properties, e.~g.~signal-to-noise ratio and size, that have been applied to both \metacal~and \im3shape, the number of  galaxies comprised in each shear catalog is significantly reduced compared to that of the full Y1 Gold catalog. Still, the number of source galaxies is unprecedented for an analysis of this kind. \metacal~consists of 35 million galaxy shape estimates, of which 26 are used in the cosmological analysis due to redshift and area cuts, and \im3shape is composed of 22 million galaxies, of which 18 are used for cosmology. The fiducial results in this paper, for instance in Sec.~\ref{sec:shearratio} and Sec.~\ref{sec:bias}, utilize \metacal~due to the higher number of galaxies included the catalog.

\subsection{Photometric redshifts for the source sample}
\label{subsec:photoz}

Galaxy redshifts in DES are estimated from $griz$ multiband photometry. The performance and accuracy of these estimates was extensively tested with Science Verfication (SV) data, using a variety of photometric redshift algorithms and matched spectroscopy from different surveys \cite{Sanchez2014,Bonnett2015}. 

The fiducial photometric redshifts used in this work are estimated with a modified version of the Bayesian Photometric Redshifts (BPZ) code \cite{Benitez2000a, photoz}. BPZ defines the mapping between color and redshift by drawing upon physical knowledge of stellar population models, galaxy evolution and empirical spectral energy distributions of galaxies at a range of redshifts.

Such photo-$z$'s are used to split our source samples into four tomographic bins by the mean of the estimated individual redshift probability density functions ($p(z)$) 
between $z=0.2$ and $z=1.3$. For \metacal\ in particular, where potential selection biases need to be corrected for (cf. \autoref{sec:metacal_responses}), this is done using photo-$z$ estimates based on \metacal\ measurements of multiband fluxes.
For both shear catalogs, the corresponding redshift distributions come from stacking random draws from the $p(z)$ and are shown in Fig.~\ref{fig:N(z)}. 
Details of this procedure are described in section 3.3 of \cite{photoz}.

The photo-$z$ calibration procedure we follow in Y1 is no longer based on spectroscopic data, since existing spectroscopic surveys are not sufficiently complete over the magnitude range of the DES Y1 source galaxies. Instead, we rely on complementary comparisons to 1) matched COSMOS high-precision photometric redshifts and 2) constraints on our redshift distributions from DES galaxy clustering cross-correlations. We refer the reader to the four dedicated redshift papers \cite{photoz,xcorr,xcorrtechnique,redmagicpz}. In addition, in this work we will provide further independent validation of their calibration, using weak gravitational lensing (Sec.~\ref{sec:shearratio}).

\subsection{Lognormal simulations}
\label{subsec:sims}
Lognormal models of cosmological fields, such as matter density and cosmic shear, have been shown to accurately describe two-point statistics such as galaxy-galaxy lensing
on sufficiently large scales. 
Furthermore, the production of lognormal mock catalogs that reproduce properties of our sample is significantly less demanding in terms of computational expenses than $N$-body simulations such as those detailed in \cite{Buzzard2017}. One of the first descriptions of lognormal fields in cosmological analyses was outlined in \cite{Coles1991}. The assumption of lognormality for these cosmological fields has shown good agreement with $N$-body simulations and real data up to nonlinear scales \cite{Kayo2001,Lahav2004,Hilbert2011}. Thus, lognormal mock simulations provide a way to assess properties of the galaxy-galaxy lensing covariance matrix that are particularly dependent on the number of simulations produced, due to their low-cost nature of production.

We use the publicly available code \flask \footnote{\texttt{http://www.astro.iag.usp.br/$\sim$flask/}} \cite{Xavier2016}, to generate galaxy position and convergence fields consistent with our lens and source samples, and produce 150 full-sky shear and density mock catalogs. The maps are pixelated on a \healpix~ grid with resolution set by an $N_\mathrm{side}$ parameter of 4096. At this $N_\mathrm{side}$, the typical pixel area is 0.73~arcmin$^2$ and the maximum multipoles resolved for clustering and shear are $\ell=8192$ and $\ell = 4096$, respectively. We mask out regions of the grid to then produce eight DES Y1 footprints for a given full-sky mock. This produces a total of 1200 mock surveys that mimic our sample.

To correctly capture the covariance properties of this sample, such as shot noise, we match the number density of the mock tomographic bins to those of the data. We add noise properties to the shear fields according to the same procedure detailed in \cite{shearcorr}. Galaxy bias is introduced in the lens samples through the input angular auto and cross power spectra between bins, and is also chosen to 
approximately match the data. The tracer density fields are subsequently Poisson sampled to yield discrete galaxy positions. 

\section{Measurement and covariance}
\label{sec:measurement}

\subsection{Measurement methodology}

Here we describe the details of the tangential shear measurement $\left< \gamma_t \right>$. Similarly, we can measure the cross-component of the shear $\left< \gamma_\times \right>$, which is a useful test of possible systematic errors in the measurement as it is not produced by gravitational lensing. For a given lens-source galaxy pair $j$ we define the tangential ($e_t$) and cross ($e_\times$) components of the ellipticity of the source galaxy as
\begin{equation} \label{eq: tangential and cross ellipticity}
e_{t, j} = -\text{Re} \left[ e_j \text{e} ^{-2i\phi_j} \right] \quad , \quad e_{\times, j}= -\text{Im} \left[ e_j \text{e} ^{-2i\phi_j} \right] ,
\end{equation}
where $e_j = e_{1,j} + i\,e_{2, j}$, with $e_{1,j}$ and $e_{2, j}$ being the two components of the ellipticity of the source galaxy measured with respect to a Cartesian coordinate system centered on the lens, and $\phi_j$ being the position angle of the source galaxy with respect to the horizontal axis of the Cartesian coordinate system. 
Assuming the intrinsic ellipticities of individual source galaxies are randomly aligned, we can obtain the mean weak lensing shear $\left<\gamma_{t/\times}\right>$ averaging the ellipticity measurements for each component over many such lens-source pairs. However, note that the assumption of random galaxy orientations is broken by intrinsic galaxy alignments (IA), which lead to non-lensing shape correlations (e.g. \cite{Troxel2015}), which are included in the modelling of the combined probes cosmology analysis \cite{keypaper}). Then:
\begin{equation}\label{eq: gamma_t estimation}
\left<\gamma_{\alpha} \left(\theta\right) \right>= \frac{\sum_j \omega_j e_{\alpha, j}  }{\sum_j \omega_j},
\end{equation} 
where $\theta$ is the angular separation, $\alpha=t$ or $\times$ denotes the two possible components of the shear and $w_j = w_{\rm l} \, w_{\rm s} \, w_{\rm e}$ is a weight associated with each lens-source pair, which will depend on the lens ($w_{\rm l}$, see \ref{subsec:observing_conditions}), on the source weight assigned by the shear catalog ($w_{\rm s}$, see \ref{sec:metacal_responses} \& \ref{sec:im3shape_responses}) and on a weight assigned by the estimator ($w_{\rm e}$, see App.~\ref{sec:deltasigma}). These estimates need to be corrected for shear responsivity (in the case of \metacal~shears, \ref{sec:metacal_responses}) or multiplicative and additive bias (in the case of \im3shape, \ref{sec:im3shape_responses}). Also note that in this work $w_{\rm e} = 1$ because we are using the $\gamma_t$ estimator, which weights all sources uniformly. Another option would be to choose an optimal weighting scheme that takes into account the redshift estimate of the source galaxies to maximize the lensing efficiency, as it is the case of the $\Delta \Sigma$ estimator. In the context of a cosmological analysis combining galaxy-galaxy lensing and cosmic shear, using uniform weighting for the sources has the considerable advantage that nuisance parameters describing the systematic uncertainty of shear and redshift estimates of the sources are the same for both probes. In Appendix~\ref{sec:deltasigma}, we find the increase in signal-to-noise ratio due to the optimal weighting scheme to be small given the photo-$z$ precision of source galaxies in DES, and hence we use the $\gamma_t$ estimator in this work to minimize the number of nuisance parameters in the DES Y1 cosmological analysis \cite{keypaper}.

In all measurements in this work, we grouped the galaxy pairs in 20 log-spaced angular separation bins between 2.5 and 250 arcmin. We use \texttt{TreeCorr}\footnote{\texttt{https://github.com/rmjarvis/TreeCorr}} \cite{Jarvis2004} to compute all galaxy-galaxy lensing measurements in this work.

One advantage of galaxy-shear cross-correlation over
shear-shear correlations is that additive shear systematics (with
constant $\gamma_1$ or $\gamma_2$) average to zero in the tangential coordinate system. However, this cancellation only occurs when sources are distributed isotropically around the lens 
and additive shear is spatially constant, 
two assumptions that are not accurate in practice, especially 
near the survey edge or in heavily masked regions, where there is a lack of symmetry on the source distribution around the lens. To remove additive systematics robustly, we also measure the tangential shear around random points: such points have no net lensing signal (see Sec.~\ref{subsec:cross-component}), yet they sample the survey edge and masked regions in the same way as the lenses. Our full estimator of tangential shear can then be written as:
\begin{equation} \label{eq: random points subtraction}
\left< \gamma_\alpha (\theta) \right> = \left< \gamma_\alpha (\theta) _{\text{Lens}} \right>-  \left< \gamma_\alpha (\theta) _{\text{Random}}\right>.
 \end{equation} 
Besides accounting for additive shear systematics, removing the measurement around random points from the measurement around the lenses has other benefits, such as leading to a significant decrease of the uncertainty on large scales, as was studied in detail in \cite{Singh2016}. We further discuss the implications the random point subtraction has on our measurement and covariance in App.~\ref{sec:appendix_rp}.

\subsubsection{\textsc{Metacalibration} responses}
\label{sec:metacal_responses}

\begin{figure*}
	\includegraphics[width=0.9\textwidth]{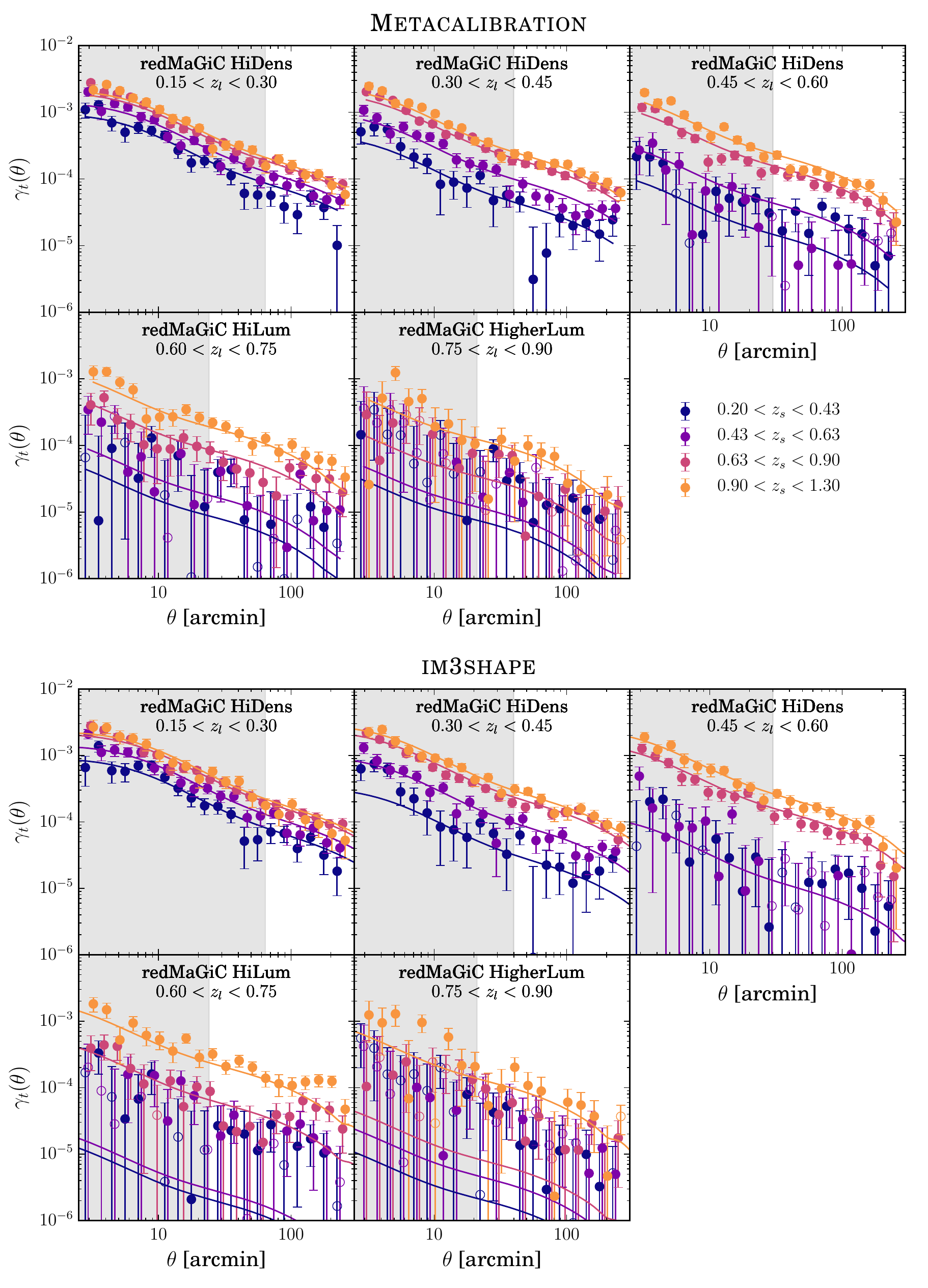}
    \caption{Tangential shear measurements for \textsc{Metacalibration} and \textsc{im3shape} together with the best-fit theory lines from the DES Y1 multiprobe cosmological analysis \cite{keypaper}. Scales discarded for the cosmological analysis, smaller than $12 \, h^{-1} \mathrm{Mpc}$ in comoving distance, but which are used for the shear-ratio test, are shown as shaded regions. Unfilled points correspond to negative values in the tangential shear measurement, which are mostly present in the lens-source combinations with low signal-to-noise due to the lenses being at higher redshift than the majority of sources.  \texttt{HiDens}, \texttt{HiLum} and \texttt{HigherLum} correspond to the three redmagic samples (High Density, High Luminosity and Higher Luminosity) described in Sec.~\ref{subsec:redmagic}}.
    \label{fig:measurement}
\end{figure*}

In the \metacal~shear catalog \cite{Huff2017,Sheldon2017,shearcat}, shears are calibrated using the measured response of the shear estimator to shear, which is usually the ellipticity $\boldsymbol{e}=(e1,e2)$. Expanding this estimator in a Taylor series about zero shear
\begin{equation} \label{eq:metacal}
\begin{split}
\boldsymbol{e} &= \left. \boldsymbol{e}\right|_{\gamma=0} + \left.  \frac{\partial \boldsymbol{e}}{\partial \boldsymbol{\boldsymbol{\gamma}}}\right|_{\gamma=0} \boldsymbol{\gamma} + ... \\
&\equiv  \left. \boldsymbol{e}\right|_{\gamma=0} +  \boldsymbol{R}_\gamma \boldsymbol{\gamma} + ... \, ,
\end{split}
\end{equation}
we can define the shear response $\boldsymbol{R}_\gamma$, which can be measured for each galaxy by artificially shearing the images and remeasuring the ellipticity:
\begin{equation}
R_{\gamma,i,j} = \frac{e_i^+ - e^{-}_i}{\Delta \gamma_j},
\end{equation}
where $e_i^+$, $e_i^-$ are the measurements made on an image sheared by $+\gamma_j$ and $-\gamma_j$, respectively, and $\Delta \gamma_j=2\gamma_j$. In the Y1 \metacal\ catalog, $\gamma_j = 0.01$. If the estimator $\boldsymbol{e}$ is unbiased, the mean response matrix $\left<R_{\gamma,i,j}\right>$ will be equal to the identity matrix.

Then, averaging  Eq.~(\ref{eq:metacal}) over a sample of galaxies and assuming the intrinsic ellipticities of galaxies are randomly oriented, we can express the mean shear as:
\begin{equation}
\left< \gamma\right> \approx \left< \boldsymbol{R}_\gamma \right> ^{-1}  \left< \boldsymbol{e} \right> 
\end{equation}
It is important to note that any shear statistic will be effectively weighted by the same responses. Therefore, such weighting needs to be included when averaging over quantities associated with the source sample, for instance when estimating redshift distributions (cf. \cite{photoz}, their section 3.3). We are including these weights in all the redshift distributions measured on \metacal~used in this work.

Besides the shear response correction described above, in the \metacal~framework, when making a selection on the original catalog using a quantity that could modify the distribution of ellipticities, for instance a cut in S/N, it is possible to correct for selection effects. In this work, we are taking this into account when cutting on S/N and size (used in Sec.~\ref{subsec:size_snr} to test for systematics effects) and in BPZ photo-$z$'s (used to construct the source redshift tomographic bins). This is performed by measuring the mean response of the estimator to the selection, repeating the selections on quantities measured on sheared images. Following on the example of the mean shear, the mean selection response matrix $\left<\boldsymbol{R}_S \right>$  is
\begin{equation}
\left<R_{S,i,j} \right>  = \frac{\left< e_i\right>^{S_+}-\left< e_i\right>^{S_-}}{\Delta \gamma_j},
\end{equation}
where $\left<e_i\right>^{S_+}$ represents the mean of ellipticities measured on images without applied shearing
in component $j$, 
but with selection based on parameters from positively sheared images. $\left<e_i\right>^{S_-}$ is the analogue quantity for negatively sheared images. 
In the absence of selection biases, $\left<\boldsymbol{R}_S \right>$ would be zero. Otherwise, the full response is given by the sum of the shear and selection response: 
\begin{equation}\label{eq:responses_sum}
\left<\boldsymbol{R} \right> =  \left<\boldsymbol{R}_\gamma \right> + \left<\boldsymbol{R}_S \right>.
\end{equation}
The application of the response corrections depends on the shear statistic that is being calibrated; a generic correction for the two point functions, including the tangential shear, which is our particular case of interest, is derived in \cite{Sheldon2017}. In this work we make use of two approximations that significantly simplify the calculation of the shear responses. First, in principle we should take the average in Eq.~(\ref{eq:responses_sum}) over the sources used in each bin of $\theta$, but we find no significant variation with $\theta$ and use a constant value (see App.~\ref{sec:appendix_responses}). Therefore, the correction to the tangential shear becomes just the average response over the ensemble. Second, we assume the correction to be independent of the relative orientation of galaxies, so that we do not rotate the response matrix as we do with the shears in Eq.~(\ref{eq: tangential and cross ellipticity}). Overall, our simplified estimator of the tangential shear for \metacal, which replaces the previous expression from Eq.~(\ref{eq: gamma_t estimation}) is:
\begin{equation}
\left<\gamma_{t,\text{mcal}}\right>= \frac{1}{\left< R_\gamma\right> + \left<R_S\right> }\frac{\sum_j \omega_{\mathrm{l},j} \, e_{t,j}}{\sum_j \omega_{\mathrm{l},j}},
\end{equation}
summing over lens-source or random-source pairs $j$ and where $\omega_{\mathrm{l},j} $ are the weights associated with the lenses.

The measured selection effects due to sample selection and photo-$z$ binning for each tomographic bin are $0.0072$, $0.014$, $0.0098$ and $0.014$, which represent 0.99\%, 2.1\%, 1.5\% and 2.4\% of the total response in each bin.

\subsubsection{\textsc{im3shape} calibration }
\label{sec:im3shape_responses}
For the \textsc{im3shape} shear catalog, additive and multiplicative corrections need to be implemented in the following manner, replacing the previous expression from Eq.~(\ref{eq: gamma_t estimation}) \cite{shearcat}:

\begin{equation}
\left< \gamma_{t,\mathrm{im3shape}} \right>= \frac{\sum_j \omega_{\mathrm{l},j}\, \omega_{\mathrm{s},j}    \, e_{t, j} }{\sum_j   \omega_{\mathrm{l},j}\, \omega_{\mathrm{s},j}  \, (1+m_j)},
\end{equation}
summing over lens-source or random-source pairs $j$, where $m_j$ is the multiplicative correction and the additive correction $c_j$ has to be applied to the Cartesian components of the ellipticity, before the rotation to the tangential component, defined in Eq.~(\ref{eq: tangential and cross ellipticity}), has been performed. $\omega_{\mathrm{l},j} $ are the weights associated with the lenses and $\omega_{\mathrm{s},j} $ the ones associated with the \im3shape catalog.
\vspace{0.4cm}

From here on, we will refer to the mean tangential shear $\left< \gamma_t\right>$  as $\gamma_t$ for simplicity.

\subsection{Measurement results}

We present the DES Y1 galaxy-galaxy lensing measurements in Fig.~\ref{fig:measurement}. The total detection significance using all angular scales for the fiducial \metacal\ catalog corresponds to $S/N = 73$. \redd{Signal-to-noise is computed as in \cite{shearcorr}, $S/N = (\gamma_t^{\mathrm{data}}C^{-1}\gamma_t^{\mathrm{model}})/(\sqrt{\gamma_t^{\mathrm{data}}C^{-1}\gamma_t^{\mathrm{model}}})$, where $C$ and $\gamma_t^{\mathrm{model}}$ are the covariance matrix and the best-fit models for galaxy-galaxy lensing measurements in the DES Y1 cosmological analysis \cite{keypaper}}. A series of companion papers present other two-point functions of galaxies and shear on the same data sample, as well as the associated cosmological parameter constraints from the combination of all these two-point function measurements \cite{wthetapaper,shearcorr,keypaper}. The shaded regions from this figure correspond to scales that are excluded in the multiprobe cosmological analysis, i.e., scales smaller than $12\, h^{-1} \mathrm{Mpc}$ in comoving distance for the galaxy-galaxy lensing observable \cite{methodpaper}. In the top panel we present the measurements for the \metacal~shear catalog, and for \im3shape in the bottom panel. Note that the measurements from the two shear catalogs cannot be directly compared, since their populations and thus their corresponding redshift distributions differ. For each of the five lens redshift bins, we measure the tangential shear for four tomographic source bins, which result in 20 lens-source redshift bin combinations. The relative strength of the galaxy-galaxy lensing signal for a given lens bin depends on the geometry of the lens-source configuration. This feature is exploited in the shear-ratio test, presented in Sec.~\ref{sec:shearratio}, where we constrain the mean of the source redshift distributions using the small scales that are not used in the cosmological analysis (shaded in Fig~\ref{fig:measurement}).

\subsection{Covariance matrix validation}\label{subsec:cov_matrix}

\begin{figure*}
	\includegraphics[width=0.9\textwidth]{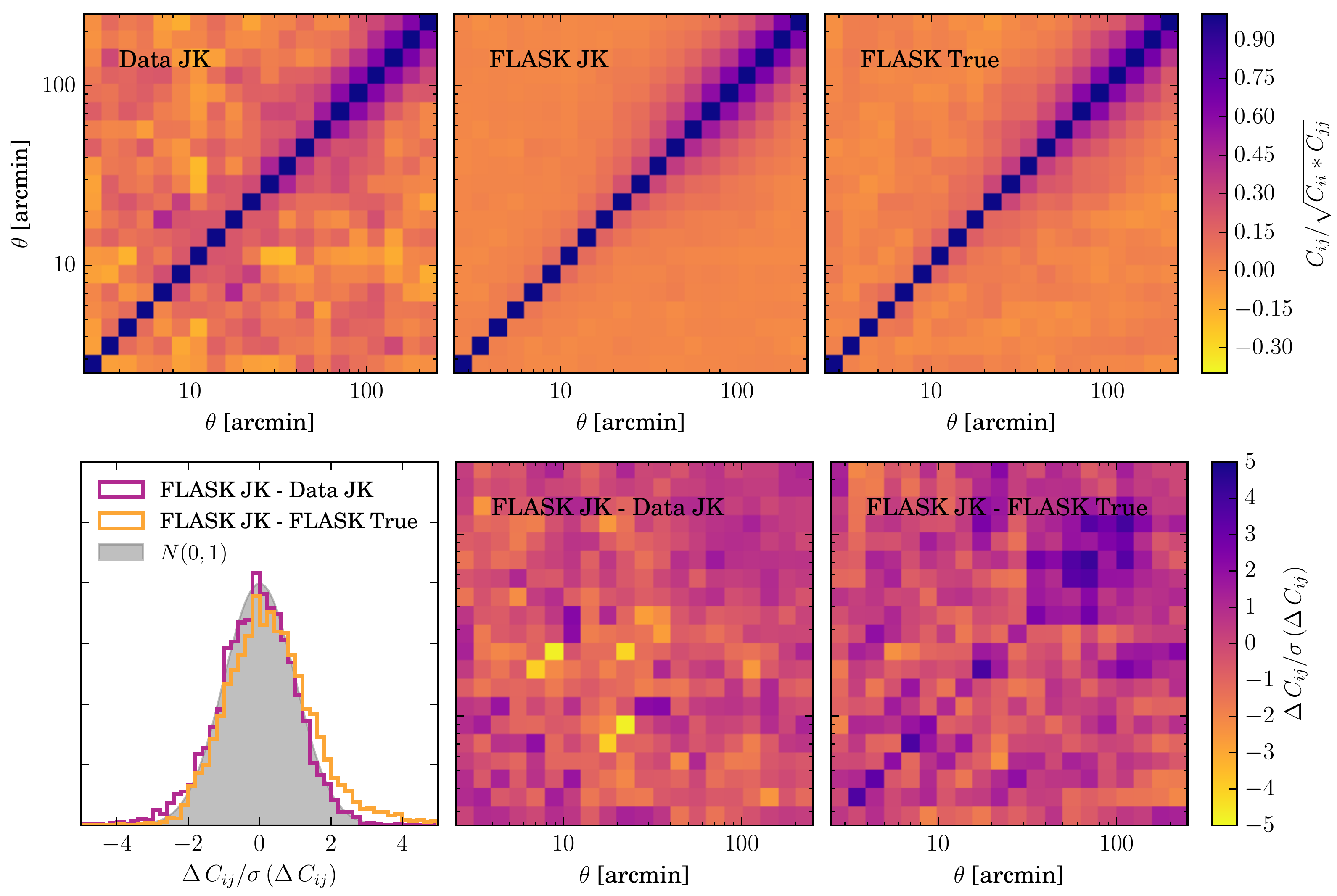}
    \caption{Correlation matrices obtained from the jackknife method on the data (top-left panel), from the mean of jackknife covariances using 100 FLASK realizations (top-middle panel) and from the 1200 lognormal simulations FLASK (top-right panel), for an example redshift bin ($0.3<z_l<0.45$ and $0.63<z_s<0.90$). In the bottom-middle and bottom-right panels, we show the differences between the covariance matrices shown in the upper panels normalized by the uncertainty on the difference, for the same example redshift bin. On the bottom-left panel, we display the normalized histograms of these  differences ($20\times20$ for each covariance, corresponding to 20 angular bins) for all the $5\times 4$ lens-source redshift bin combinations, compared to a Gaussian distribution centered at zero with a width of one.}
    \label{fig:cov_sims_comparison}
\end{figure*}

\begin{figure*}
	\includegraphics[width=\textwidth]{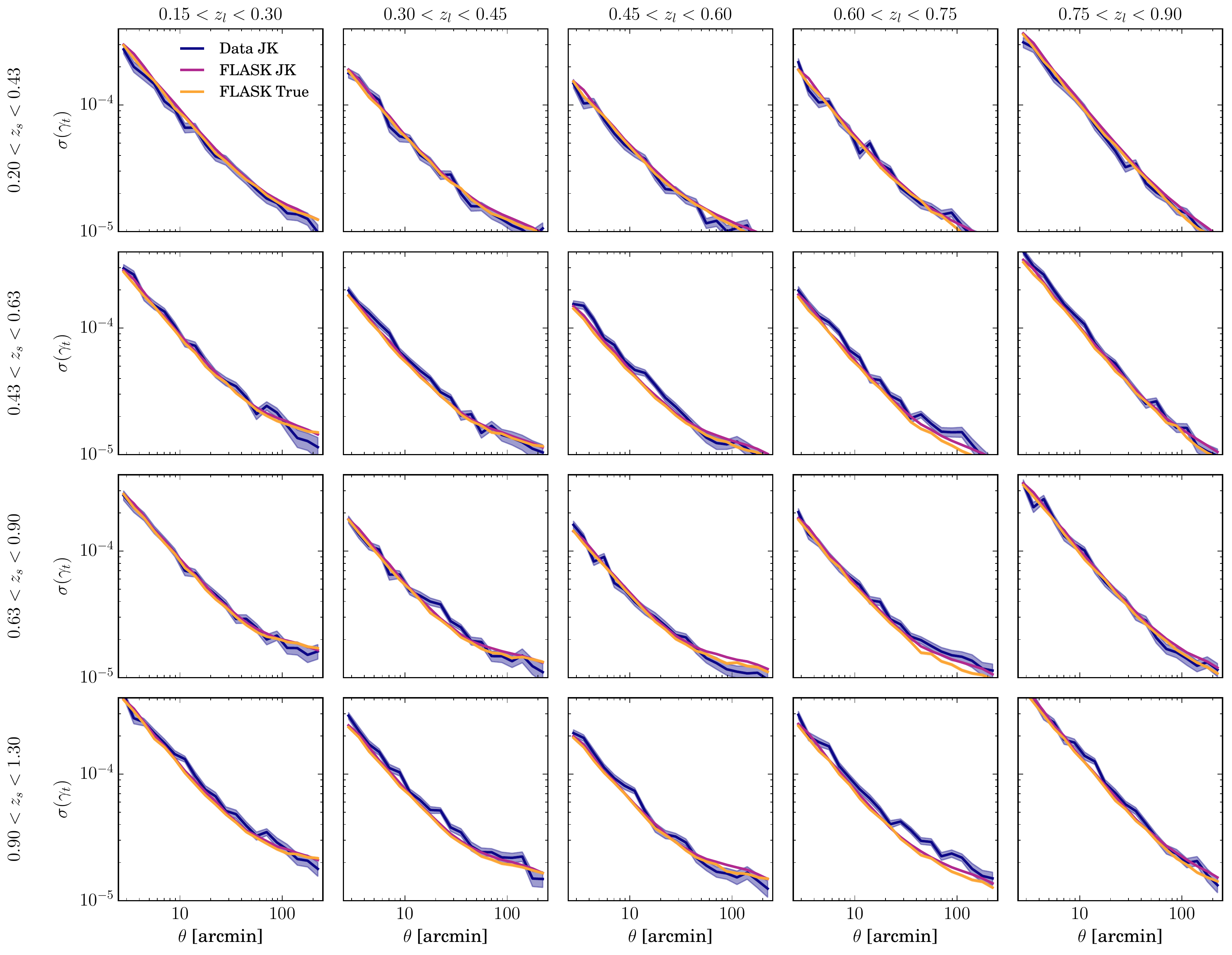}
    \caption{Comparison of the diagonal elements of the covariance obtained from the jackknife method on the data (\texttt{Data JK}), from the mean of jackknife covariances using 100 FLASK realizations (\texttt{FLASK JK}) and from the 1200 lognormal simulations FLASK (\texttt{FLASK True}), for all the lens-source combinations.}
    \label{fig:sigma_gammat_sims_comparison}
\end{figure*}

Galaxy-galaxy lensing measurements are generally correlated across angular bins. The correct estimation of the covariance matrix is crucial not only in the usage of these measurements for cosmological studies but also in the assessment of potential systematic effects that may contaminate the signal. While a validated halo-model covariance is used for the DES Y1 multiprobe cosmological analysis \cite{methodpaper}, in this work we use jackknife (JK) covariance matrices given the requirements of some systematics tests performed here, such as splits in area, size or S/N. A set of 1200 lognormal simulations, described in Section~\ref{subsec:sims}, is used to validate the jackknife approach in the estimation of the galaxy-galaxy lensing covariances.  We estimate the JK covariance using the following expression:
\begin{equation}
C_{ij}^{\text{JK}}\, (\gamma_{i}, \gamma_{j})= \frac{N_{\rm JK}-1}{N_{\rm JK}} \, \sum_{k=1}^{N_{\rm JK}} \, \left( \, \gamma_{i}^k - \overline{\gamma_{i}}\,\right) \, \left( \, \gamma_{j}^k - \overline{\gamma_{j}}\, \right),
\end{equation} 
where the complete sample is split into a total of $N_{\mathrm{JK}}$ regions, $\gamma_{i}$ represents either $\gamma_t (\theta_i)$ or $\gamma_\times (\theta_i)$, $\gamma_{i}^k$ denotes the measurement from the $k^{th}$ realization and the $i^{th}$ angular bin, and  $\overline{\gamma_{i}}$ is the mean of $N_{\rm JK}$ resamplings.

Jackknife regions are obtained using the \texttt{kmeans} algorithm\footnote{\texttt{https://github.com/esheldon/kmeans\_radec}}
run on a homogeneous random point catalog with the same survey geometry and, then, all foreground catalogs
(lenses and random points) are split in $N_{\mathrm{JK}}=100$ 
subsamples. Specifically, \texttt{kmeans} is a clustering algorithm that subdivides $n$ objects into $N$ groups (see Appendix B in \cite{Suchyta2016} for further details).

In the upper panels of Fig.~\ref{fig:cov_sims_comparison} we present the different covariance estimates considered in this work, namely the jackknife covariance in the data (\texttt{Data JK}), the mean of 100 jackknife covariances measured on the lognormal simulations (\texttt{FLASK JK}) and the true covariance from 1200 lognormal simulations (\texttt{FLASK True}), for a given lens-source redshift bin combination ($0.3<z_l<0.45$ and $0.63<z_s<0.90$). On the lower panels of this figure, we show the differences between them normalized by the corresponding uncertainty. The lower left panel shows the distribution of these differences and its agreement with a normal distribution with $\mu=0$ and $\sigma=1$, as expected from a pure noise contribution, using all possible lens-source bin combinations, and the lower middle and right panels show the same quantity element-by-element for the redshift bin combination used in the upper panels. The uncertainty on the data jackknife covariance comes from the standard deviation of the jackknife covariances measured on 100 lognormal simulations. The uncertainties on the two other covariance estimates are significantly smaller; in the mean of 100 jackknife covariances it is $\sqrt{N}$ times smaller, where $N=100$ in our case. On the other hand, the uncertainty on each element of the true covariance from 1200 lognormal simulations is calculated using $(\Delta C_{ij})^2 = (C_{ii} C_{jj}  + C_{ij} C_{ij})/(N-1)$, where $N=1200$ in our case. The lower left panel shows an overall good agreement between the covariance estimates, even though the larger tail of the orange histogram with respect to a normal distribution indicates a potential slight overestimation of the covariance obtained with the jackknife method.

In Fig.~\ref{fig:sigma_gammat_sims_comparison} we compare the diagonal elements of the covariance for the 20 lens-source redshift bin combinations, obtaining good agreement for all cases and scales.  As in Fig.~\ref{fig:cov_sims_comparison}, the uncertainty on the data jackknife covariance comes from the standard deviation of the jackknife covariances measured on 100 lognormal simulations. The uncertainties on the two other error estimates are also shown on the plot, but are of the same order or smaller than the width of the lines. 

Overall, we have validated the implementation of the jackknife method on the data by comparing this covariance to the application of the same method on 100 lognormal simulations and to the true covariance obtained from 1200 lognormal simulations, and finding good agreement among them, both for the diagonal and off-diagonal elements.

\section{Data systematics tests}
\label{sec:tests}

\begin{figure}
	\includegraphics[width=\columnwidth]{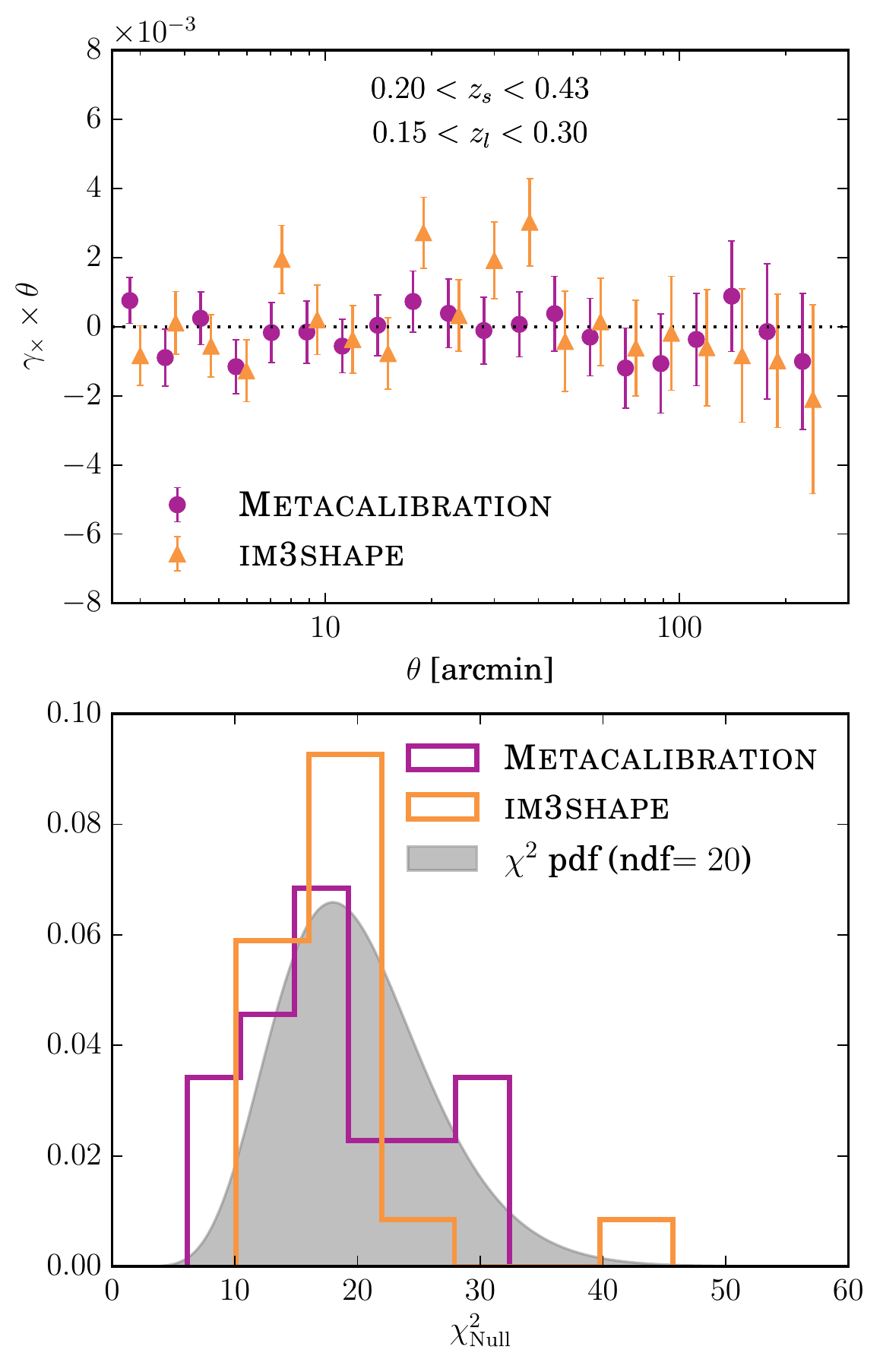}
    \caption{(\textit{Top panel}): Cross-component of the galaxy-galaxy lensing signal with random points subtraction for one lens-source redshift bin combination. (\textit{Bottom panel}): The null $\chi^2$ histogram from all $5\times 4$ lens-source redshift bins combinations computed with the jackknife covariance corrected with the Hartlap factor \cite{Hartlap2007}, compared to the $\chi^2$ distribution with 20 degrees of freedom corresponding to 20 angular bins. We find the cross-component to be consistent with zero. }
    \label{fig:cross-component}
\end{figure}

In order to fully exploit the power of weak gravitational lensing, we need to measure the shapes of millions of tiny, faint galaxies to exceptional accuracy, and possible biases may arise from observational, hardware and software systematic effects. Fortunately, weak lensing provides us with observables that are very sensitive to cosmology and the physical properties of the objects involved but also with others for which we expect no cosmological signal. By measuring such observables, we can characterize and correct for systematic effects in the data. In this section, we perform a series of tests that should produce a null signal when applied to true gravitational shear, but whose non-zero measurement, if significant, would be an indication of systematic errors leaking into the galaxy-galaxy lensing observable.

\subsection{Cross-component}
\label{subsec:cross-component} 

The mean cross-component of the shear $\gamma_\times$, which is rotated 45 degrees with respect to the tangential shear and is defined in Eq.~(\ref{eq: tangential and cross ellipticity}), should be compatible with zero if the shear is only produced by gravitational lensing, since the tangential shear captures all the galaxy-galaxy lensing signal. Note that the cross-component would also be null in the presence of a systematic error that is invariant under parity. 

In the top panel of Fig.~\ref{fig:cross-component} we show the resulting cross-shear measured around redMaGiC lenses (including random point subtraction) for one lens-source redshift bin combination and for both shear catalogs. In the bottom panel we display the null $\chi^2$ histogram coming from all $5\times 4$ lens-source $\gamma_\times$ measurements, computed using the jackknife covariance for the cross-component, described and validated in Sec.~\ref{subsec:cov_matrix}. To compute the null $\chi^2$, i.e. $\chi^2_\text{null} = \gamma_\times^T \, C^{-1} \,  \gamma_\times$, we need an estimate of the inverse of the covariance matrix, but since jackknife covariance matrices contain a non-negligible level of noise, we need to correct for the fact that the inverse of an unbiased but noisy estimate of the covariance matrix is not an unbiased estimator of the inverse of the covariance matrix \cite{Hartlap2007}. Thus, we apply the Hartlap correcting factor $(N_{\rm JK} - p -2)/(N_{\rm JK}-1)$ to the inverse covariance, where $N_{\rm JK}$ is the number of jackknife regions and $p$ the number of angular bins. Our results indicate the cross-component is consistent with zero.

\subsection{Impact of PSF residuals}
\label{subsec:psf_residuals}

The estimation of source galaxy shapes involves modeling them convolved with the PSF pattern, which depends on the atmosphere and the telescope optics and which we characterize using stars in our sample. Next, we test the impact of residuals in the PSF modeling on the galaxy-galaxy lensing estimator, and we compare the size of this error to the actual cosmological signal.

\begin{figure}
	\includegraphics[width=\columnwidth]{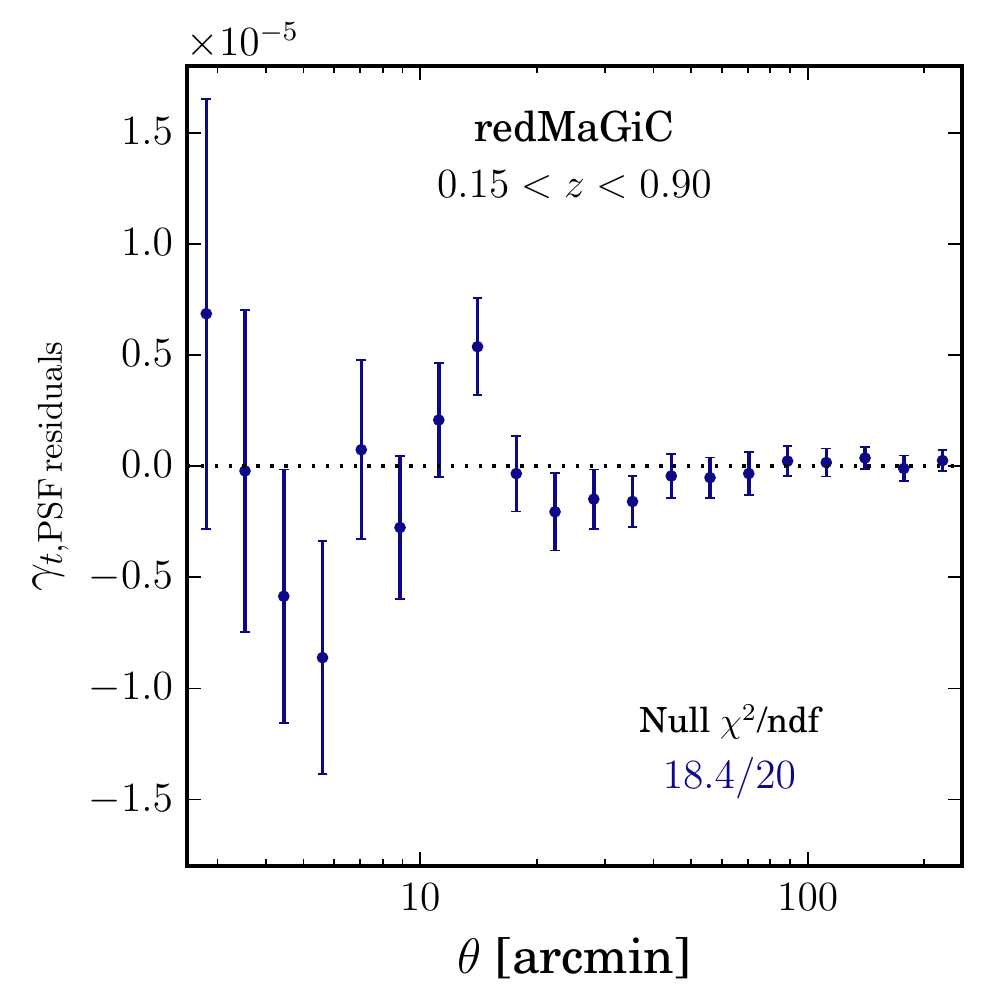}
    \caption{PSF residuals for \texttt{PSFEx} model, using a single non-tomographic lens bin, including random-point subtraction. It is consistent with a null measurement and much smaller than the signal.}
    \label{fig:psfresiduals}
\end{figure}

Explicitly, the PSF residuals are the differences between the measured shape of the stars and the \texttt{PSFEx} model \cite{Bertin2011,shearcat} at those same locations. In Fig.~\ref{fig:psfresiduals} we show the measured mean of the tangential component of the PSF residuals around redMaGiC galaxies, including the subtraction of the same quantity around random points, in the same manner as for the tangential shear signal. We find it is consistent with zero, and also much smaller than the signal 
(cf.~\autoref{fig:measurement}).

\begin{figure*}
	\includegraphics[width=0.8\textwidth]{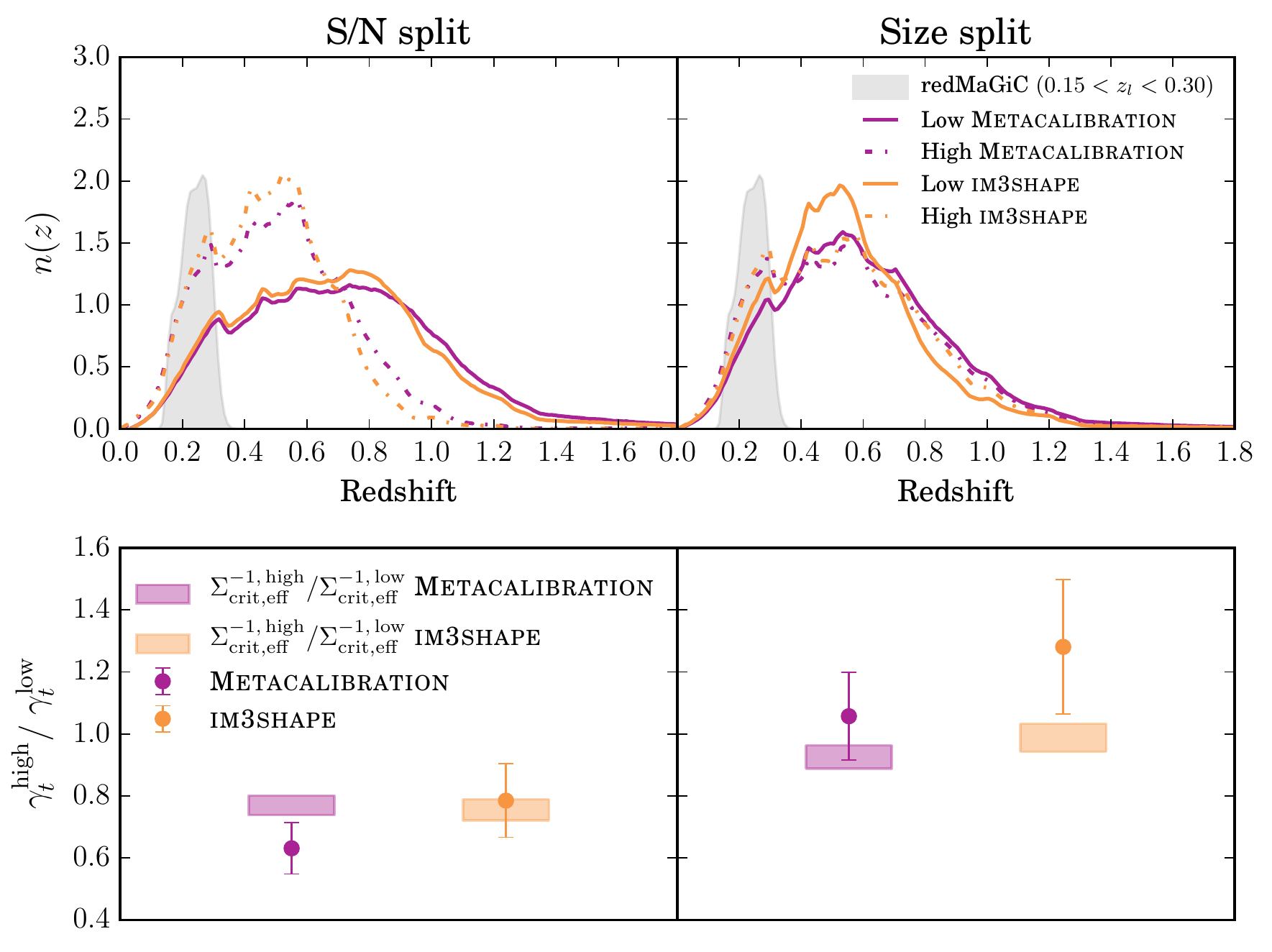}
    \caption{S/N (\textit{left}) and size (\textit{right}) splits tests for \textsc{Metacalibration} and \textsc{im3shape}, using scales employed in the cosmology analysis ($>12 \, h^{-1} \mathrm{Mpc}$). (\textit{Top panels}): Redshift distributions of the lens and source samples used for this test. (\textit{Bottom panels}): Comparison between the ratio of $\Sigma_{\mathrm{crit},\mathrm{eff}}^{-1}$ using the above redshift distributions (\textit{boxes}) to the ratio between the amplitudes coming from the fit of the tangential shear measurement for each half to the smooth template from the lognormal simulations (\textit{points}).}
    \label{fig:snr_size}
\end{figure*}

\subsection{Size and S/N splits}\label{subsec:size_snr}

Potential biases in shape measurements are likely to be more important for galaxies which are either small or detected at low signal-to-noise (S/N). Even though the shape measurement codes utilized in this work are calibrated in a way such that these effects are taken into account, it is important to test for any residual biases in that calibration. In order to perform such a test, we split the source galaxy samples in halves of either low or high size or S/N, and examine the differences between the galaxy-galaxy lensing measurements using the different halves of the source galaxy samples. For this test, we use the lower redshift lens bin to minimize the overlap in redshift with the source samples. The sources are all combined into a single bin, to maximize the sensitivity to potential differences between the halves. 

In order to estimate the size of galaxies, for \metacal~we use \textit{round} measure of size (\texttt{T\_r}), and for \im3shape we use the $R_{\rm gpp}/R_{\rm p}$ size parameter, both  defined in \cite{Jarvis2015}. We estimate the S/N of galaxies using the \textit{round} measure of S/N for \metacal, (\texttt{s2n\_r}), and the \texttt{snr} quantity for \im3shape, both  defined in \cite{Jarvis2015}. Splitting the source galaxy samples in halves of low and high galaxy S/N or size, we measure the corresponding galaxy-galaxy lensing signals, and we check their consistency.

Since these quantities can correlate with redshift, differences can arise in the redshift distributions between the halves of S/N and size splits, as seen in the upper panels of Fig.~\ref{fig:snr_size}. When comparing the tangential shear signals of each half of the split, we therefore need to account for the differences in the lensing efficiency given by the two redshift distributions. We do this in the following way. From Eq.~(\ref{eq:gammat_delta_sigma}), the ratio between the tangential shear measurements for each half of the split in the absence of systematics effects is
\begin{equation}\label{eq:sigma_crit_eff}
 \frac{\gamma_{t}^{l,s_\mathrm{high}}}{\gamma_{t}^{l, s_\mathrm{low}}} =
\frac{\Sigma_{\mathrm{crit},\mathrm{eff}}^{-1\ l,s_\mathrm{high}}}{\Sigma_{\mathrm{crit},\mathrm{eff}}^{-1\ l,s_\mathrm{low}}},
 \end{equation} 
since $\gamma_{t}^{l,s_\mathrm{high}}$ and $\gamma_{t}^{l, s_\mathrm{low}}$ share the same lens sample and thus the same $\Delta \Sigma$. $\Sigma_{\mathrm{crit},\mathrm{eff}}^{-1}$, defined in Eq.~(\ref{eq:eff_inverse_sigma_crit}), is a double integral over the lens and source redshift distributions and the geometrical factor $\Sigma_{\mathrm{crit}}^{-1}$, which depends on the distance to the lenses, the sources and the relative distance between them. Then, to check the consistency between the tangential shear measurements for each half of the source split we will compare the ratio between them to the ratio between the corresponding $\Sigma_{\mathrm{crit},\mathrm{eff}}^{-1}$'s.

Then, the validity of this test to flag potential biases in shape measurements related to S/N and size is linked to an accurate characterization of the redshift distributions. The ensemble redshift distributions are estimated by stacking the redshift probability density functions of individual galaxies in each split, as given by the BPZ photo-$z$ code. As described in \cite{photoz} and a series of companion papers \cite{xcorrtechnique,xcorr,redmagicpz} we do not rely on these estimated redshift distributions to be accurate, but rather calibrate their expectation values using two independent methods: a matched sample with high-precision photometric redshifts from COSMOS, and the clustering of lensing sources with redMaGiC galaxies of well-constrained redshift. These offsets to the BPZ estimate of the ensemble mean redshift, however, could well be different for the two halves of each of the splits.

To estimate these calibration differences between the subsamples, we repeat the COSMOS calibration of the redshift distributions (see \cite{photoz} for details), splitting the matched COSMOS samples by \metacal\ size and signal-to-noise ratio at the same thresholds as in our data. We find that the shifts required to match the mean redshifts of the subsamples with the mean redshifts of the matched COSMOS galaxies are different by up to $|\Delta(\Delta z)|=0.035$ for the overall source sample. 

In the upper panels of Fig.~\ref{fig:snr_size}, the mean values of the redshift distributions have been corrected using the results found in the analysis described above, and these corrected $n(z)$'s are the ones that have been used in the calculation of $\Sigma_{\mathrm{crit},\mathrm{eff}}^{-1}$ in Eq.~(\ref{eq:sigma_crit_eff}). The ratio of $\Sigma_{\mathrm{crit},\mathrm{eff}}^{-1}$'s is shown in the lower panels of Fig.~\ref{fig:snr_size} and its uncertainty comes from the propagation of the error in the mean of the source redshift distributions for each half of the split, i.e. $\sqrt{2}$ times the non-tomographic uncertainty as estimated in \cite{photoz} using COSMOS.

Regarding the left-hand side of Eq.~(\ref{eq:sigma_crit_eff}), to avoid inducing biases from taking the ratio between two noisy quantities, we fit an amplitude for each half of the split to a smooth tangential shear measurement that we obtain from the mean of tangential shear measurements on 100 independent log-normal simulations. Then, we take the ratio between the amplitudes fitted for each half of the split. We repeat this procedure for each data jackknife resampling, obtaining a ratio for each of those, whose mean and standard deviation are shown in the lower panels of Fig.~\ref{fig:snr_size} (\textit{points}), compared to the ratio of $\Sigma_{\mathrm{crit},\mathrm{eff}}^{-1}$'s (\textit{boxes}).

Given the uncertainties in both the measurements and the photometric redshift distributions presented in Fig.~\ref{fig:snr_size}, we find no significant evidence of a difference in the galaxy-galaxy lensing signal when splitting the \metacal~or \im3shape source samples by size or S/N. Specifically, we find a 1.6$\,\sigma$ (0.24$\,\sigma$) difference for the \metacal~(\im3shape) S/N split and a 0.90$\,\sigma$ (1.3$\,\sigma$) difference for the \metacal~(\im3shape) size split.

\subsection{Impact of observing conditions}\label{subsec:observing_conditions}

Time-dependent observing conditions are intrinsic to photometric surveys, and they may impact the derived galaxy catalogs, for instance, introducing galaxy density variations across the survey footprint. In this section we test for potential biases in the galaxy-galaxy lensing measurements due to these differences in observing conditions and their effect in the survey galaxy density. We use projected \texttt{HEALPix} 
\cite{Gorski2005} sky maps (with resolution $N_{\mathrm{side}}=4096$) in the $r$ band for the following quantities:

\begin{itemize}
\item \textbf{\texttt{AIRMASS}:} Mean airmass, computed as the optical path length for light from a celestial object through Earth's atmosphere (in the secant approximation), relative to that at the zenith for the altitude of CTIO.
\item \textbf{\texttt{FWHM}:} Mean seeing, i.e., full width at half maximum of the flux profile.
\item \textbf{\texttt{MAGLIMIT}:} Mean magnitude for which galaxies are detected at $S/N = 10$. 
\item \textbf{\texttt{SKYBRITE}:} Mean sky brightness. 
\end{itemize}

More information on these maps can be found in \cite{y1gold} and \cite{wthetapaper}.

In order to test for potential systematic effects, we split each map into halves of high and low values of a given quantity, and measure the galaxy-galaxy lensing signal in each half. We are using the same configuration as in the S/N and size splits, i.e. the lower redshift lens bin and a single non-tomographic source bin between $0.2 < z_s <1.3$. In this case, we are splitting both the lens and the source samples, since the split is performed in area. 

\begin{figure}
	\includegraphics[width=\columnwidth]{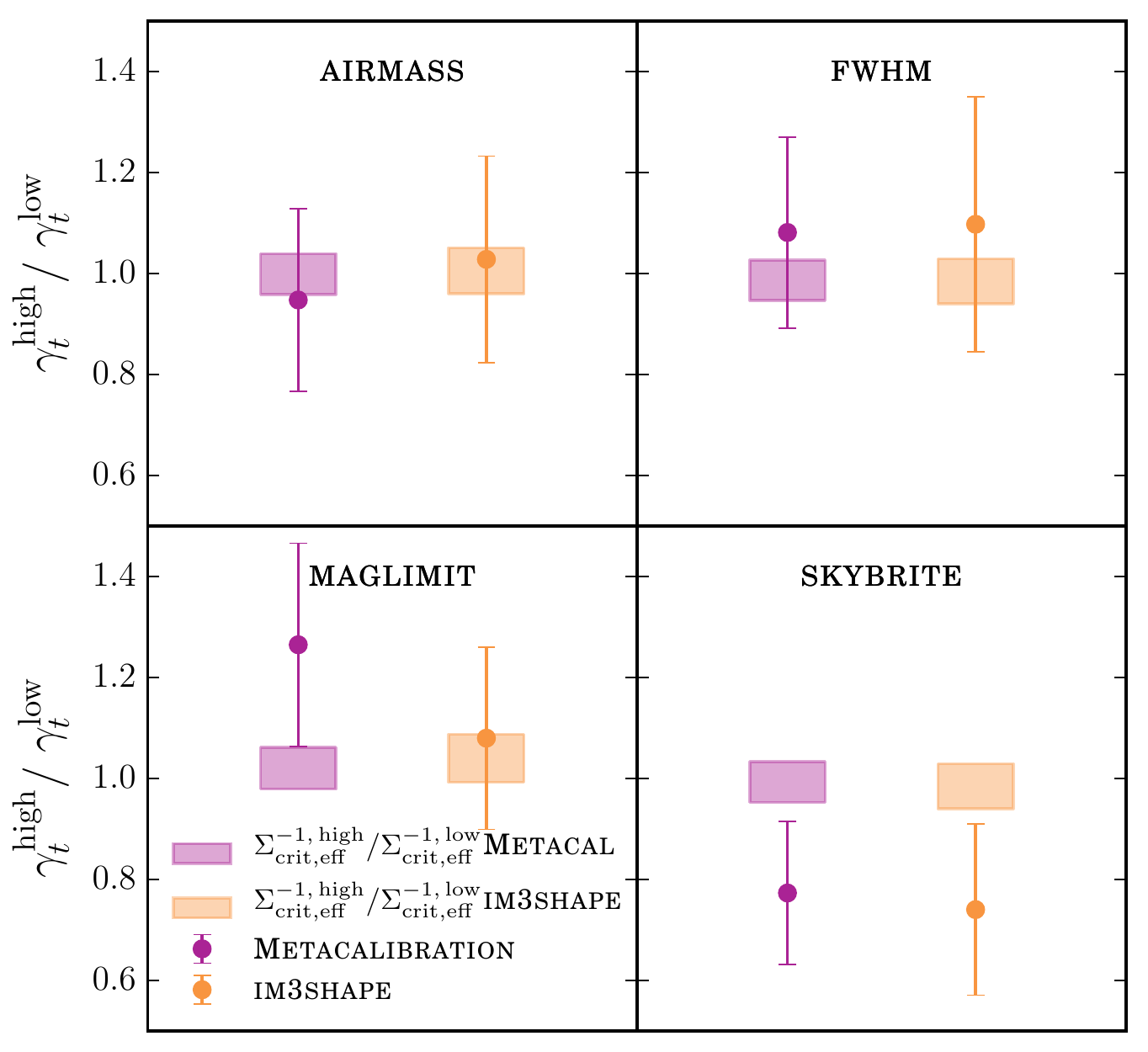}
    \caption{Results for the tests involving area splits in halves of different observational systematics maps in the $r$ band, with angular scales used in the cosmology analysis ($12 \, h^{-1} \mathrm{Mpc}$). We compare the ratio of $\Sigma_{\mathrm{crit},\mathrm{eff}}^{-1}$ (\textit{boxes}) using the redshift distributions for each split to the ratio between the amplitudes coming from the fit of the tangential shear measurement for each half to the smooth template derived from the lognormal simulations (\textit{points}), following the same procedure as for the S/N and size splits, described in Sec.~\ref{subsec:size_snr} and shown in Fig.~\ref{fig:snr_size}.}
    \label{fig:systematics_maps}
\end{figure}

To check the consistency between the measurements in each half we follow the same approach as for the S/N and size splits, described in detail in the previous section, where we take into account the differences in the redshift distributions of the sources. 
We find the correlation between observing conditions and redshift to be very mild for the source sample, as can be seen in Fig.~\ref{fig:systematics_maps}, where the ratios of $\Sigma_{\mathrm{crit},\mathrm{eff}}^{-1}$'s are all compatible with unity.  For the lens sample this correlation is even smaller, consistent with the lens sample containing brighter and lower-redshift galaxies. The differences on the mean redshift between the lens redshift distributions of the two halves are of the order of $0.001$ or smaller for all maps, which is negligible for this test, although we have not performed independent calibration of redshift biases for these split samples. 

The results for these area splits are shown in Fig.~\ref{fig:systematics_maps} for \metacal~and \im3shape. In most cases, the ratio between the measurements on each half of the splits lie within 1$\sigma$ of the corresponding ratio of $\Sigma_{\mathrm{crit},\mathrm{eff}}^{-1}$'s, and at slightly more than 1$\sigma$ in the remaining cases. Thus, we do not encounter any significant biases on the galaxy-galaxy lensing signal due to differences in observing conditions.

The effect of the same variable observing conditions in the galaxy clustering measurements using the same DES redMaGiC sample is studied in detail in \cite{wthetapaper}. In that analysis, maps which significantly correlate with galaxy density are first identified, and then a set of weights is computed and applied to the galaxy sample so that such dependency is removed, following a method similar to that presented in \cite{Ross2012, Ross2017}. The resulting set of weights from that analysis has been also used in this work, for consistency in the combination of two-point correlation functions for the DES Y1 cosmological analysis. Nonetheless, the impact of such a weighting scheme in the galaxy-galaxy lensing observables is found to be insignificant, consistent with the tests presented above in this section and with previous studies (see \cite{Kwan2016}).     

\begin{figure*}
	\includegraphics[width=0.88\textwidth]{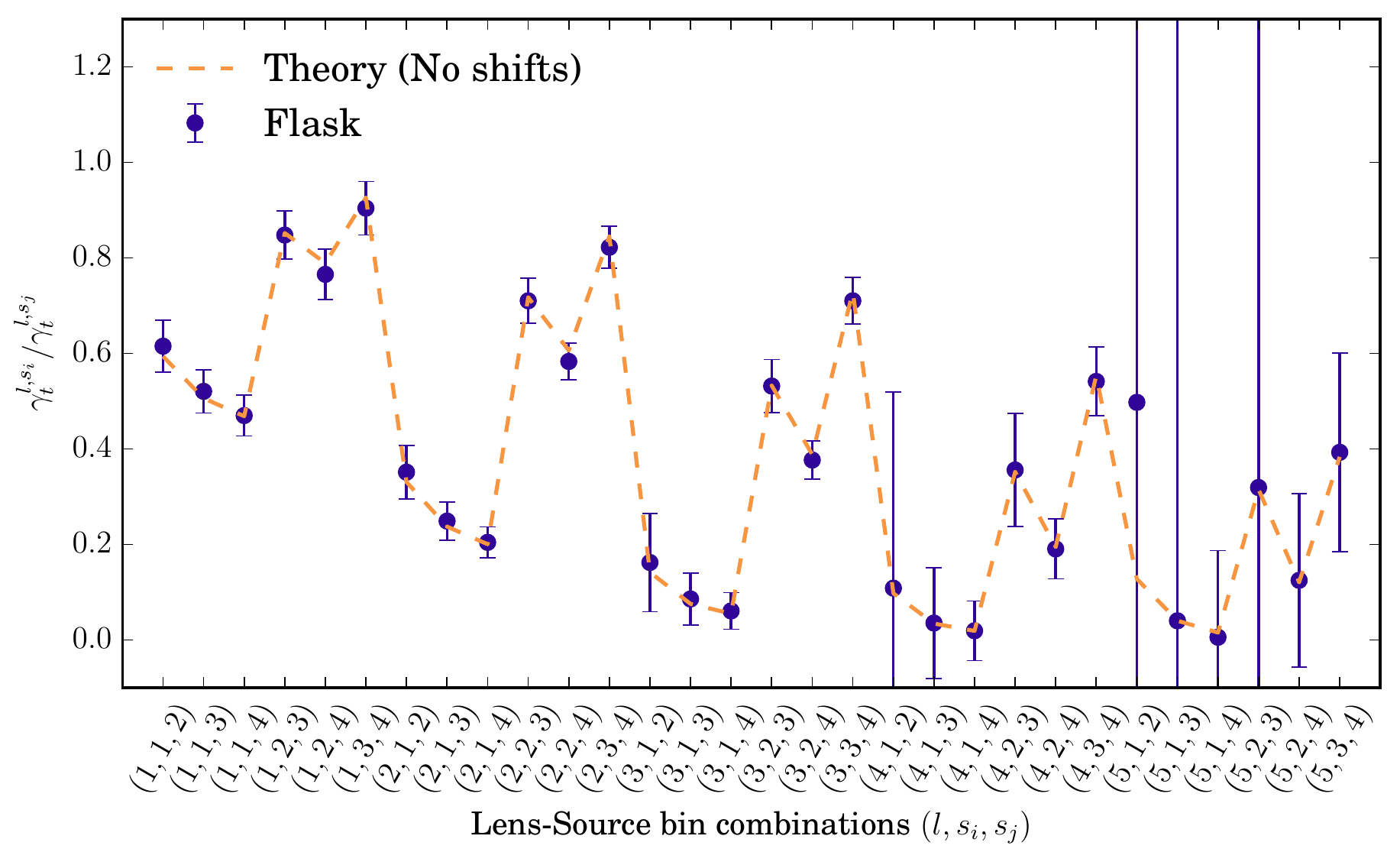}
    \caption{Comparison between the mean ratio of tangential shear measurements using 1200 independent log-normal simulations and the ones calculated from theory, for all lens-source bin ratio combinations sharing the same lens bin. The errorbars correspond to the standard deviation of the measurement on individual simulations, thus being representative of the errors that we will obtain from the data. }
    \label{fig:shear-ratio-vector-flask}
\end{figure*}

\begin{figure*}
	\includegraphics[width=0.88\textwidth]{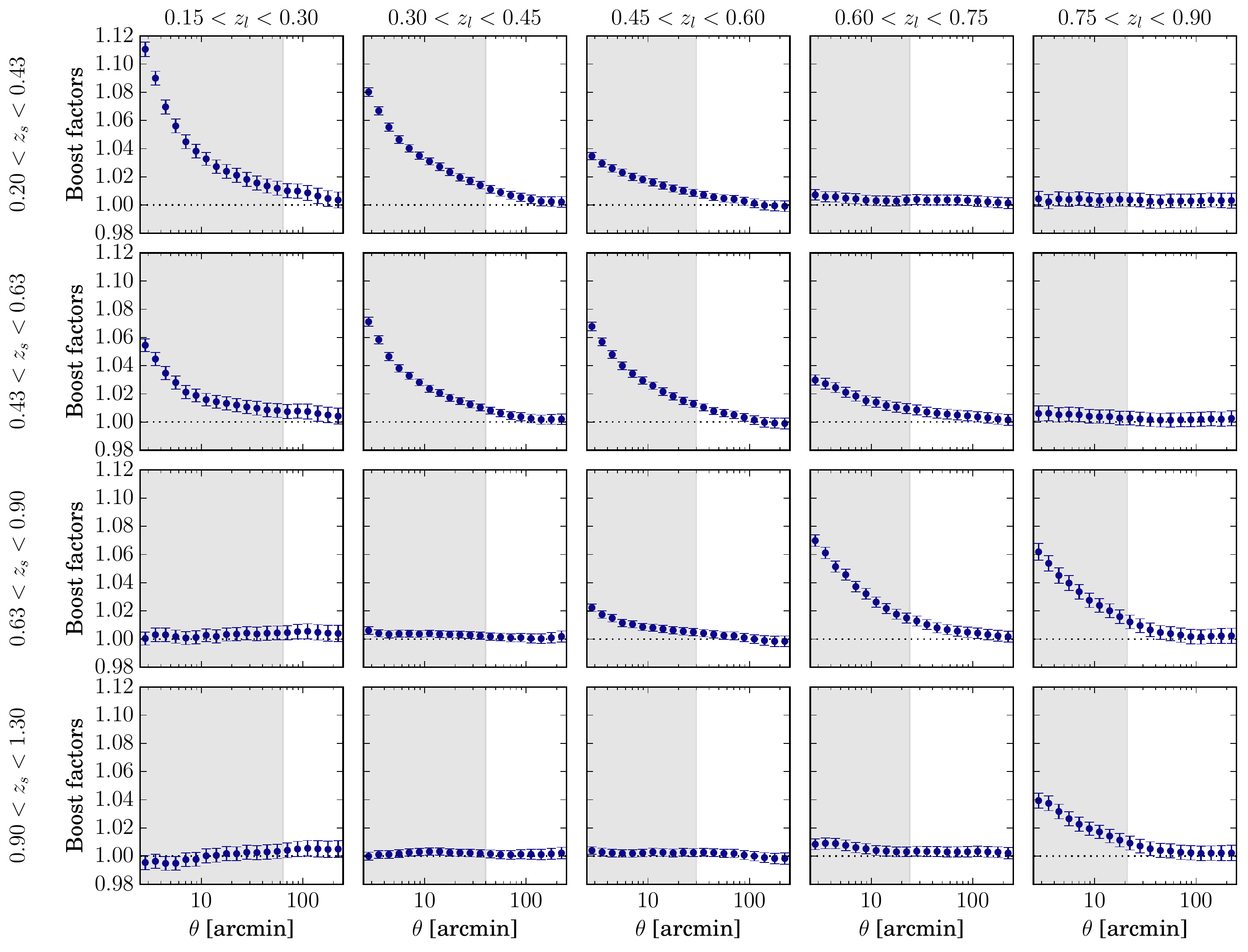}
    \caption{Boost factor correction accounting for clustering between lenses and sources in each lens-source bin used in this analysis. Non-shaded scales correspond to scales used in the DES Y1 cosmological analysis, while shaded regions are used for the shear-ratio geometrical test in this Section. Boost factors are unity or percent-level for the former, but can be significantly larger for the latter in some cases, and hence they are applied in our analysis of the shear-ratio test.}
    \label{fig:boost_factors}
\end{figure*}

\section{Shear-Ratio test} \label{sec:shearratio}

In previous sections we have seen that the variation of the galaxy-galaxy lensing signal with source redshift depends solely on the angular diameter distances relative to foreground and background galaxy populations. Such dependency was initially proposed as a probe for dark energy evolution in \cite{Jain2003}. The shear-ratio is, however, a weak function of cosmological parameters, and more sensitive to errors in the assignment of source or lens redshifts \cite{Kuijken2015}. Since redshift assignment is a crucial but difficult aspect of robust cosmological estimate for a photometric survey like DES, the shear-ratio test is a valuable cross-check on redshift assignment. In the context of the DES Y1 cosmological analysis, the usage of high-quality photometric redshifts for lens galaxies allows us to put constraints on the mean redshift of source galaxy distributions.

In this section we present a general method to constrain potential shifts on redshift distributions using the combination of ratios of galaxy-galaxy lensing measurements. First, we present the details of the implementation, and we test it on lognormal simulations. Then, we use the galaxy-galaxy lensing measurements shown in Fig.~\ref{fig:measurement}, restricted to angular scales which are not used in the DES Y1 cosmological analysis, to place independent constraints on the mean of the source redshift distributions shown in the lower panel of Fig.~\ref{fig:N(z)}. Finally, we compare our findings with those obtained from a photometric redshift analysis in the COSMOS field and from galaxy angular cross-correlations.

The ratio of two galaxy-galaxy lensing measurements around the same lens bin, hence having equivalent $\Delta\Sigma$, can be derived from Eq.~(\ref{eq:gammat_delta_sigma}) and is given by: 
\begin{equation}
\frac{\gamma_{t}^{l,s_i}}{\gamma_{t}^{l,s_j}} = \frac{\Sigma_{\mathrm{crit},\mathrm{eff}}^{-1 \ l,s_i}}{\Sigma_{\mathrm{crit},\mathrm{eff}}^{-1 \ l, s_j}},
\label{eq:shearratio1}
\end{equation} 
where $\Sigma_{\mathrm{crit},\mathrm{eff}}^{-1}$ is the double integral over lens and source redshift distributions defined in Eq.~(\ref{eq:inverse_sigma_crit}). Therefore, for two given $\gamma_t$ measurements sharing the same lens population but using two different source bins, we can predict their ratio from theory by using the estimated redshift distributions involved. In addition, we can allow for a shift in each of those redshift distributions and use the $\gamma_t$ measurements to place constraints on them.

In this section we generalize this approach by including all possible combinations of ratios of galaxy-galaxy lensing measurements sharing a given lens bin, and allowing for independent shifts in their redshift distributions. With the purpose of providing constraints on the shifts of redshift distributions which are independent of the measurement involved in the fiducial DES Y1 cosmological analysis, we restrict the galaxy-galaxy lensing measurements used for this shear-ratio test to scales smaller than the ones used by the cosmological analysis but which have still been tested against systematic effects in this work.

In order to estimate the ratio of galaxy-galaxy lensing measurements, which can be noisy and thus bias their ratio, we fit each measurement involved in the ratio, both around the same lens bin, to a power law fit of the highest signal-to-noise $\gamma_t$ measurement for the same lens bin. That fixes the shape of the galaxy-galaxy lensing signal around that lens galaxy sample. Then, fits to the amplitude of this power law are used to obtain the shear ratio.

\subsection{Testing the method on simulations}

With the purpose of testing our method to estimate ratios of galaxy-galaxy lensing measurements and our ability to recover the expected values from theory, we use the lognormal simulations described in Sec.~\ref{subsec:sims}, where we know the true lens and source redshift distributions. For that case, we should be able to find good agreement between measurements and theory, without the necessity of allowing for any shifts in the redshift distributions. 

Figure \ref{fig:shear-ratio-vector-flask} shows all the possible ratios of two $\gamma_t$ measurements in the FLASK simulations sharing the same lens bin using the lens-source binning configuration used throughout this paper (as depicted in Fig.~\ref{fig:N(z)}), with the error bars coming from the variance of the 1200 simulations. It also shows the expected values for the ratios given from theory, using the true corresponding redshift distributions with no shifts applied. The agreement between measurements and theory is excellent, demonstrating that the method described in this section is able to recover the true values of $\gamma_t$ measurements from theory when the redshift distributions are known.

\subsection{Application to data}

Now we turn to data, and utilize this shear-ratio method to constrain possible biases in the mean of redshift distributions. The lens and source redshift bins considered and their fiducial estimated redshift distributions are depicted in Figure~\ref{fig:N(z)}. The high-precision photometric redshifts of the redMaGiC sample ensure the lens redshift distributions are well known, with potential shifts found to be very small and consistent with zero in \cite{redmagicpz}, and hence we keep them fixed. On the contrary, source galaxies are generally fainter and have a much larger uncertainty in their redshift distributions. Therefore, we allow for an independent shift $\Delta z^i$ in each of the measured source redshift distributions $n_{\mathrm{obs}}^i(z)$, such that 
\begin{equation}
n_{\mathrm{pred}}^i(z) = n_{\mathrm{obs}}^i(z-\Delta z^i),
\end{equation}
to be constrained from the combination of ratios of galaxy-galaxy lensing measurements through their impact in the $\Sigma_{\mathrm{crit},\mathrm{eff}}^{-1}$ factors in Eq.~(\ref{eq:shearratio1}).

When turning to the data case, we also have to consider effects which are not included in the simulations. In particular, next we take into account the effects of potential boost factors and multiplicative shear biases in the measurements.

\subsubsection{Boost factors}\label{sec:bf}

The calculation of the mean galaxy-galaxy lensing signal in Eq.~(\ref{eq:cosmo}) correctly accounts for the fact that some source galaxies are in front of lenses due to overlapping lens and source redshift distributions, but only under the assumption that the galaxies in those distributions are homogeneously distributed across the sky. As galaxies are not homogeneously distributed but they are clustered in space, a number of sources larger than the $n^{\rm obs}(z)$ suggests may be physically associated with lenses. 
These sources are not lensed, causing a dilution of the lensing signal which can be significant at small scales. In order to estimate the importance of this effect, we compute the excess of sources around lenses compared to random points \cite{Sheldon2004}:
\begin{equation}
B(\theta) = \frac{N_{r}}{N_{l}} \frac{\sum_{l,s} w_{l,s}}{\sum_{r,s} w_{r,s}}
\label{eq:boost_factors}
\end{equation} 
where $l,s$ ($r,s$) denotes sources around lenses (random points), $w_{l,s}$ ($w_{r,s}$) is the weight for the lens-source (random-source) pair, and the sums are performed over 
an  
angular bin $\theta$. Figure \ref{fig:boost_factors} shows this calculation for every lens-source bin in this analysis. The shaded regions in the plot mark the scales used for the shear-ratio test (unused by the cosmological analysis). The importance of boost factors at small scales can be as large as 10\%, while on the large scales used for cosmology it does not depart from unity above the percent level. The data measurements used for the shear-ratio test in this section have been corrected for this effect.        

\begin{figure}
	\includegraphics[width=\columnwidth]{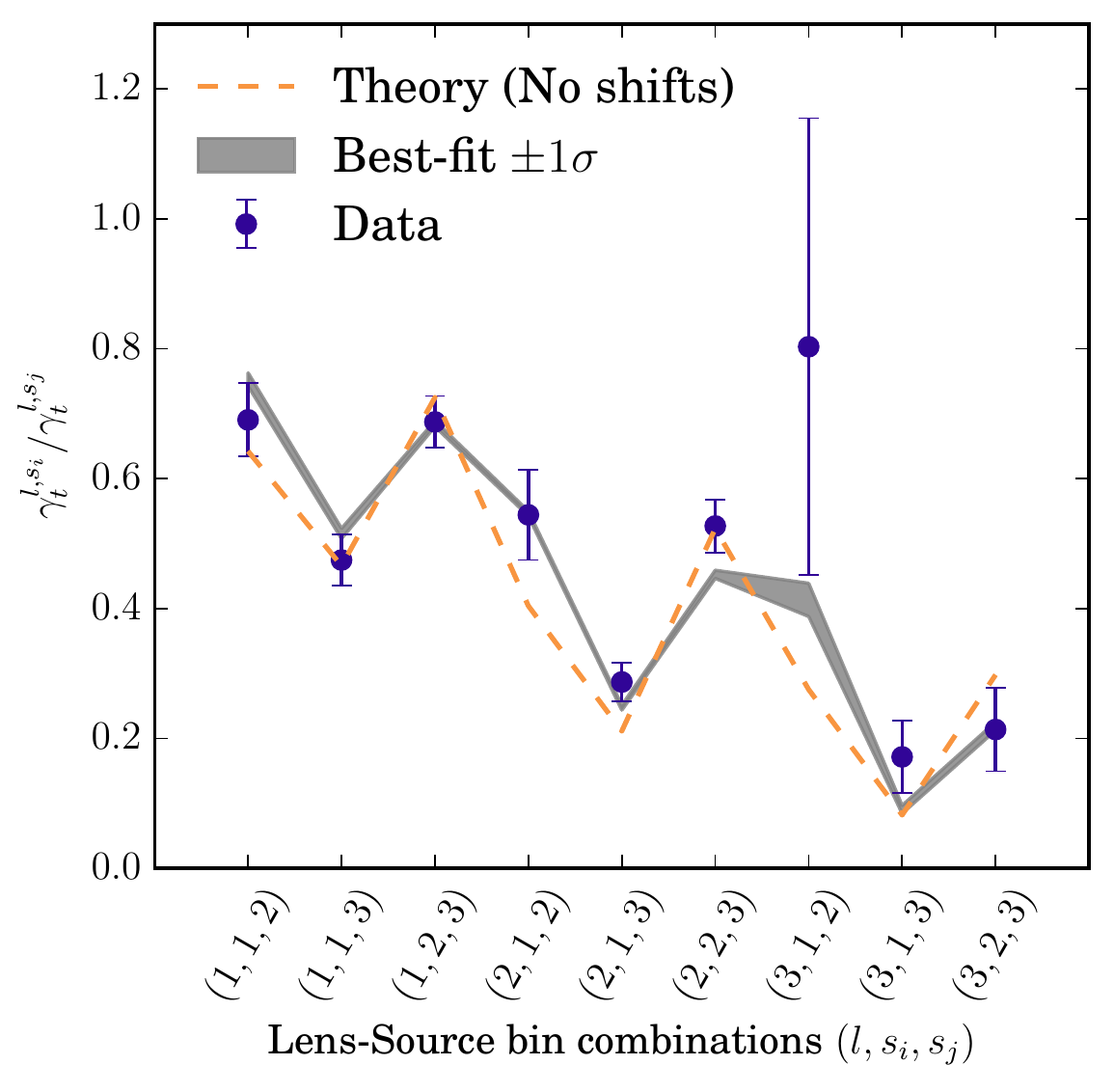}
    \caption{Comparison between the ratio of tangential shear measurements on \metacal~(\textit{blue points}) to the ones calculated from theory, both without applying any shift to the original source $n(z)'s$ (\textit{dashed orange line}) and applying the best-fit shifts with a $1\sigma$ uncertainty band (\textit{gray band}).}
    \label{fig:shear-ratio-vector-mcal}
\end{figure}

\begin{figure*}
	\includegraphics[width=0.65\textwidth]{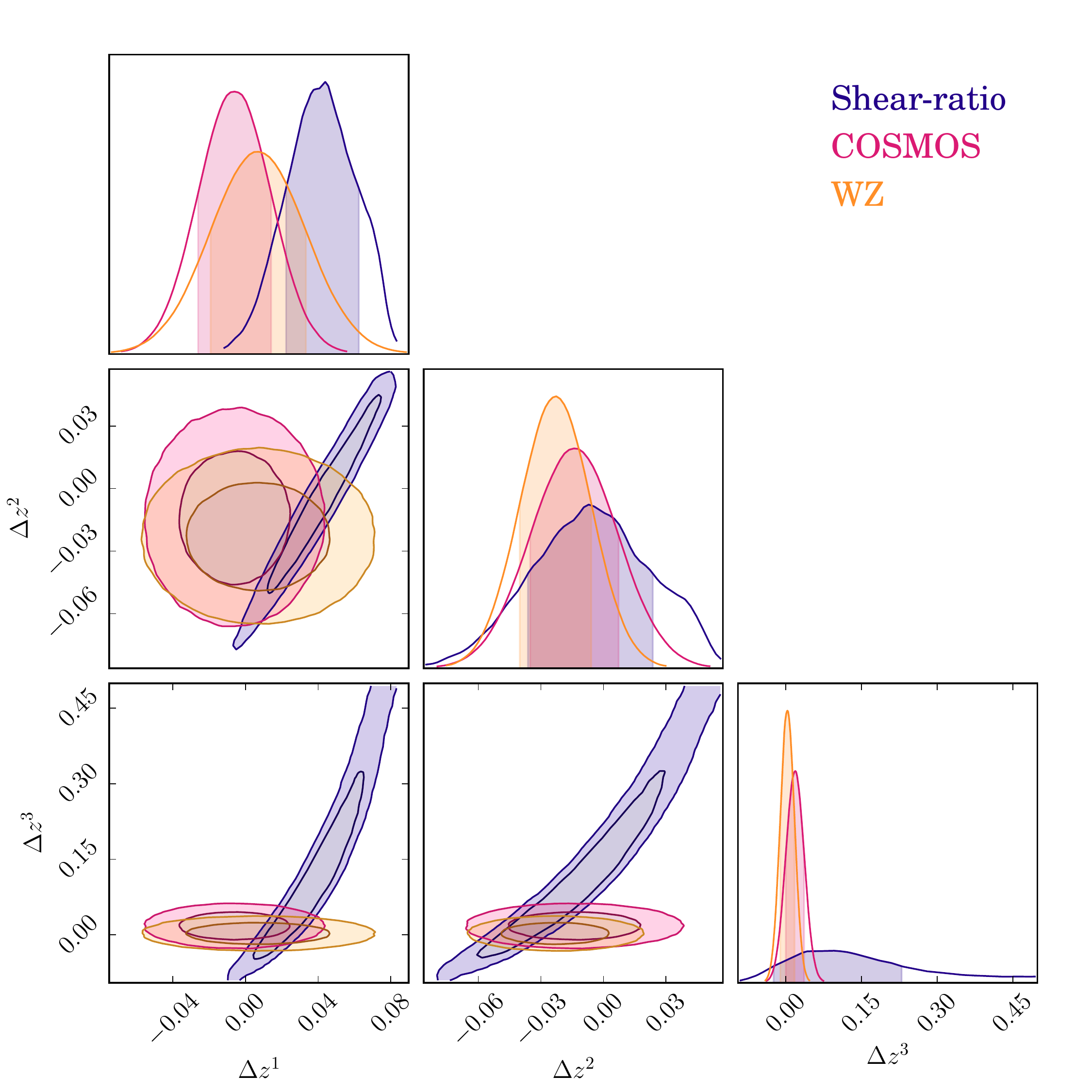}
    \caption{Comparison of the constraints obtained on the source redshift distribution shifts using different methods: Shear-ratio test, photo-$z$ studies in the COSMOS field (COSMOS, \cite{photoz}) and cross-correlation redshifts (WZ, \cite{xcorr, xcorrtechnique}).}
    \label{fig:shear-ratio-comparison}
\end{figure*}

\subsubsection{Multiplicative shear biases}

Multiplicative shear biases are expected to be present in the galaxy-galaxy lensing signal and need to be taken into account. This potential effect is included as an independent parameter $m^i$ for each source redshift bin, parametrized such that the shear ratios in Eq.~(\ref{eq:shearratio1}) look like the following: 
\begin{equation}
\frac{\gamma_{t}^{l,s_i}}{\gamma_{t}^{l,s_j}} = \frac{(1+m_j)\,\Sigma_{\mathrm{crit},\mathrm{eff}}^{-1\ l, s_i}}{(1+m^i)\,\Sigma_{\mathrm{crit},\mathrm{eff}}^{-1\ l, s_j}}.
\end{equation} 

\subsubsection{Results}

\begin{table*}[]
\setlength{\tabcolsep}{12pt}
\centering
\caption{Priors and posteriors on the mean of source redshift distributions ($\Delta z$) and multiplicative shear biases ($m$) for the first three source bins defined in this work (Fig.~\ref{fig:N(z)}), using the shear-ratio test. Priors are uniform in $\Delta z$ and Gaussian on $m$, and posteriors are given as the mean value with 68\% constraints.  }
\bigskip
\label{table:sr_results}
\begin{tabular}{c|c|c|c|c}
& $\Delta z$ Prior & $\Delta z$ Posterior & $m$ Prior & $m$ Posterior \\[0.2cm]
Source bin 1 $\;$ & Uniform($-0.5$,0.5) & $0.046^{+0.017}_{-0.023}$  & Gaussian(0.012,0.021) & $0.018^{+0.020}_{-0.021}$ \\[0.2cm] 
Source bin 2 $\;$ & Uniform($-0.5$,0.5) & $-0.005^{+0.028}_{-0.031}$ & Gaussian(0.012,0.021) & $-0.012^{+0.017}_{-0.016}$ \\[0.2cm] 
Source bin 3 $\;$ & Uniform($-0.5$,0.5) & $0.10^{+0.13}_{-0.12}$     & Gaussian(0.012,0.021) & $0.035^{+0.016}_{-0.019}$
\end{tabular}
\end{table*}

\begin{figure}
	\includegraphics[width=\columnwidth]{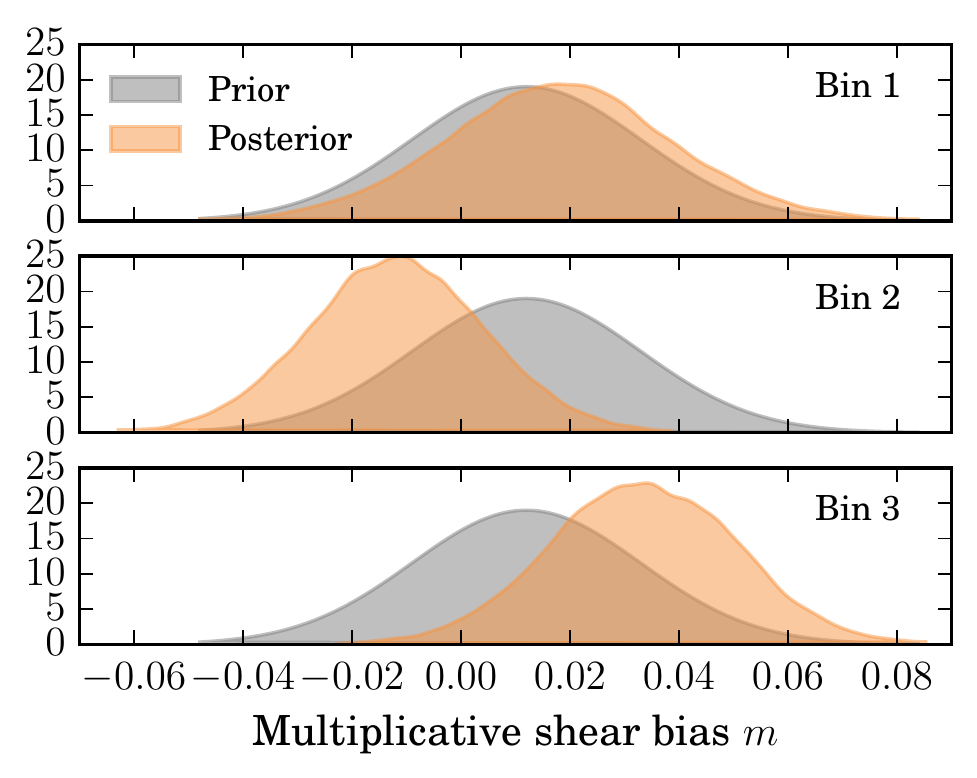}
    \caption{Prior and posterior distributions for multiplicative shear biases ($m$) for the first three source bins defined in this work (Fig.~\ref{fig:N(z)}). The shear-ratio test appears to be informative on the multiplicative shear biases for the second and third source bins, reducing the prior width by as much as 20\%, even though posteriors are all consistent with the priors at better than 1-$\sigma$ level. }
    \label{fig:shear-ratio-m}
\end{figure}

In practice, the sensitivity of the shear-ratio geometrical test to shifts in the mean of redshift distributions decreases significantly the higher the distribution is in redshift, due to the relative differences in distance with respect to the lenses and the observer being smaller for that case. For that reason, the sensitivity to shifts in the highest source redshift bin defined in this work is very small, and as there are strong correlations with the other shifts, we left out the fourth source bin. We also leave out the two highest lens redshift bins as the galaxy-galaxy lensing S/N for these cases is very small and they add little information to this test.

In order to find the best-fit shifts for all combinations of fixed-lens $\gamma_t$ ratios using these redshift bins, we set a Monte-Carlo Markov Chain (MCMC) to let the shifts vary, with a broad flat prior of [-0.5,0.5] for each shift $\Delta z^i$. We follow the recommendations in \cite{shearcat} and include a Gaussian prior of $\mu = 0.012$ and $\sigma = 0.021$ on the multiplicative shear biases $m^i$ for each source bin $i$. As the covariance is estimated from JK resampling, the corresponding Hartlap factor is applied to the covariance. \redd{Some recent studies have discussed and presented further corrections to that procedure \cite{Sellentin2016}. Given that in our case the Hartlap factor is $\simeq0.9$, such corrections would result in a small change to the parameter contours, and have not been considered in this analysis. However, a more detailed treatment of noisy covariances may need to be considered in forthcoming, more sensitive, DES analyses.}

Figure \ref{fig:shear-ratio-vector-mcal} shows the equivalent of Fig.~\ref{fig:shear-ratio-vector-flask} for the data case, including the theory prediction with no shifts and with best-fit shifts from the MCMC run, and the shear-ratio case in Fig.~\ref{fig:shear-ratio-comparison} shows the $\Delta z$ constraints from the MCMC, marginalizing over multiplicative shear biases $m$, where very clear correlations can be observed between the different shifts. In addition, Table \ref{table:sr_results} presents the derived constraints on $\Delta z$ and $m$ for the different source bins considered. Even though $\Sigma_{\mathrm{crit}}$ depends on cosmology through $\Omega_m$, the results are insensitive to that parameter to the extent that no significant changes on the shifts are observed when marginalizing over it with a broad flat prior of $0.1 < \Omega_m < 0.5$. Also, the boost factor correction from Eq.~(\ref{eq:boost_factors}) has no significant effect on the derived $\Delta z$ constraints.   

In the past, several studies have proposed shear self-calibration techniques, either from galaxy-galaxy lensing only \cite{Bernstein2006}, or using combinations of observables (e.~g.~\cite{Huterer2006,Bernstein2009}). Interestingly, the shear-ratio test can also be used as a way to calibrate potential multiplicative shear biases ($m$) present in the data. Figure \ref{fig:shear-ratio-m} displays the $m$ priors and posteriors for the three source redshift bins considered, where the posteriors show a reduction of up to 20\% in the width of the priors (see also Table \ref{table:sr_results}) for the second and third bins, therefore showing potential as a method to internally constrain shear biases in the data.      

\subsubsection{Caveats and future work}

The redshift evolution of the $\Delta\Sigma$ profile of the lens sample within a redshift bin could potentially affect the shear-ratio test and would not be noticeable in the FLASK simulations. This would especially influence the ratios between lens and source bins that are close in redshift. However, the usage of relatively thin lens tomographic bins, of 0.15 in redshift, and the little galaxy bias evolution of the redMaGiC sample for the first three lens bins, as shown in Fig.~\ref{fig:galaxy-bias} below and \cite{Clampitt2016}, suggest that this effect is small compared to our current error bars. On the other hand, mischaracterization of the tails in the fiducial (unshifted) redshift distributions of the source galaxies, especially for those close to the lenses, could also affect the results of the shifts obtained with the shear-ratio test. Studying the impact of such effects in the shear-ratio geometrical test using $N$-body simulations is beyond the scope of this paper and it is left for future work. 

In addition, intrinsic alignment (IA) between physically associated lens-source galaxy pairs can potentially affect the shear ratio measurement (see, e.~g., \cite{Sheldon2004,Blazek2012}). While IA on larger scales is modeled when measuring cosmology or the galaxy bias, we have not included this effect on the small scales used here. The boost factor measurements in Fig.~\ref{fig:boost_factors} yield an estimate of the fraction of physically associated pairs in all our measurements. As seen in \cite{Blazek2012}, for typical lensing sources the impact of IA contamination on the observed lensing signal is smaller than that of the boosts themselves. Since the boost corrections here are small and have a minimal effect on the derived source photo-$z$ shifts, we expect the impact of IA to be highly subdominant. However, it will be beneficial in future work to include the impact of IA when performing shear ratio tests.

\subsubsection{Comparison of $\Delta z$ constraints and conclusions}

In Fig.~\ref{fig:shear-ratio-comparison} we also compare the shear-ratio constraints with those obtained independently from photo-$z$ studies in the COSMOS field \cite{photoz} and from galaxy cross-correlations \cite{xcorr, xcorrtechnique}, and we find consistency among the three independent studies, with $\chi^2$/dof = 5.57/6 for the combination of the three cases. As expected, the constraining power of the shear-ratio test for the shifts on the source distributions decreases rapidly the higher the redshift of the distributions is, so that the \redd{1-D marginalized} constraints on the first tomographic bin are competitive with those from the other probes, and for the third tomographic bin they add very little information. \redd{However, on the 2-D space, the shear-ratio contours show great potential in breaking degeneracies with other probes. Therefore,} the use of this method with forthcoming data sets can have a major impact in determining possible photometric redshift biases, especially from source distributions at low redshift.

The importance of an accurate photometric redshift calibration in DES was  already noticed in the analysis of Science Verification data, where it proved to be one of the dominant systematic effects \cite{TheDarkEnergySurveyCollaboration2015}. For this reason, showing the consistency of constraints derived from galaxy-galaxy lensing only to those from more traditional photo-$z$ methods and from galaxy angular cross-correlations represents an important demonstration of the robustness of the companion DES Y1 cosmological analysis.

\section{redMaGiC galaxy bias}\label{sec:bias}

Galaxy-galaxy lensing is sensitive to cosmological parameters and the galaxy bias of the corresponding lens galaxy population, as expressed in Eq.~(\ref{eq:cosmo}). Similarly, the galaxy clustering of the same lens population also depends on both cosmology and the galaxy bias, but with a different power of the latter \cite{wthetapaper}. Therefore, the combination of galaxy clustering and galaxy-galaxy lensing breaks the degeneracy between the galaxy bias and cosmological parameters. This combination is one of the more promising avenues to understand the underlying physical mechanism behind dark energy, and has been used together with cosmic shear measurements to produce cosmological results from DES Y1 \cite{keypaper}.  

\begin{figure}[h]
	\includegraphics[width=\columnwidth]{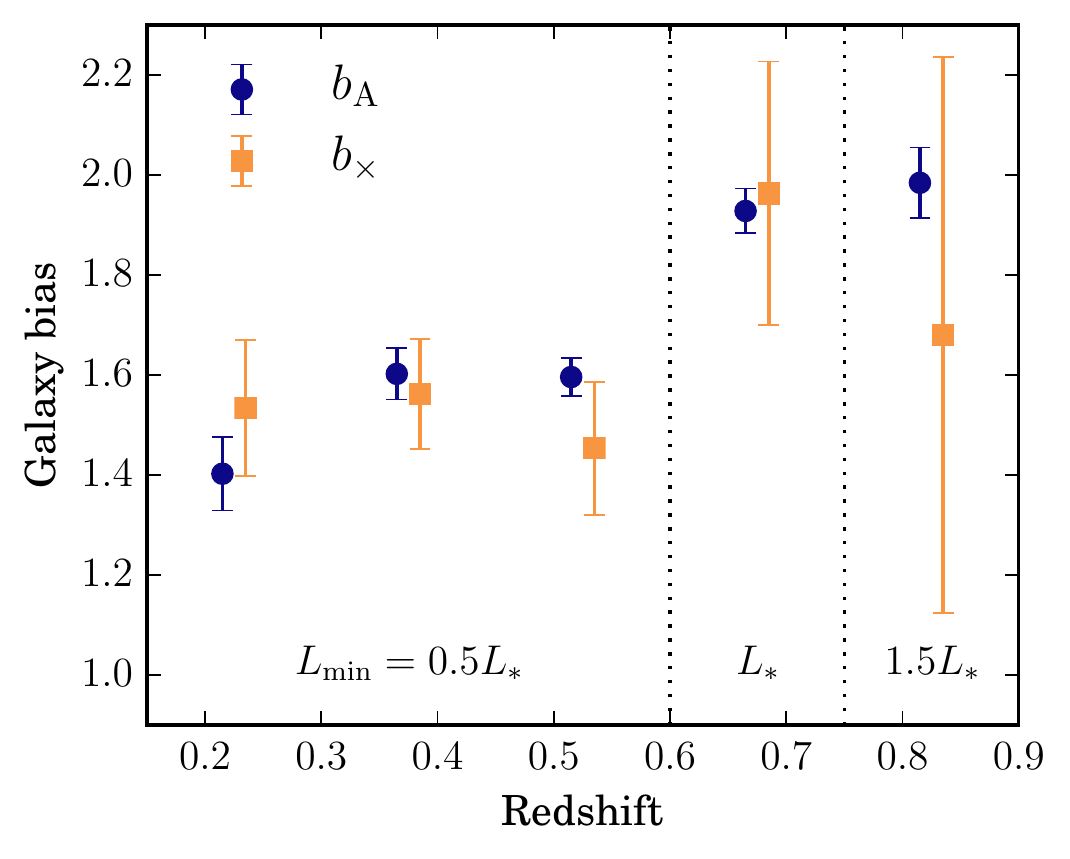}
    \caption{Comparison of the galaxy bias results obtained from galaxy clustering measurements ($b_A$, \cite{wthetapaper}) and from the galaxy-galaxy lensing measurements in this work ($b_\times$), by fixing all cosmological parameters to the 3x2 cosmology best-fit from \cite{keypaper}. The vertical dotted lines separate the three redMaGiC samples, which have different luminosity thresholds $L_\mathrm{min}$, defined in Sec.~\ref{subsec:redmagic}.}
    \label{fig:galaxy-bias}
\end{figure}

Alternatively, fixing all cosmological parameters, the measurements of galaxy clustering and galaxy-galaxy lensing can provide independent measurements of the galaxy bias of a given lens population. The DES Y1 cosmology analysis relies on the assumption that the linear bias from galaxy clustering and from galaxy-galaxy lensing is the same, which is known to break down on the small-scale regime \cite{Baldauf2010}. To verify this assumption over the scales used in the DES Y1 cosmology analysis, we measure the galaxy bias from each probe separately. In Fig.~\ref{fig:galaxy-bias} we show the bias constraints from galaxy clustering (or galaxy autocorrelations, $b_A$) and galaxy-galaxy lensing (or galaxy-shear cross-correlations, $b_\times$) on the five lens redMaGiC tomographic bins defined in this work, fixing all cosmological parameters to the best-fit obtained in the DES Y1 cosmological analysis \cite{keypaper}. We use comoving angular separations larger than 8$h^{-1}$Mpc for galaxy clustering, and larger than 12$h^{-1}$Mpc for galaxy-galaxy lensing, which correspond to the scales used in the DES Y1 cosmological analysis. In order to obtain these results, the clustering measurements from \cite{wthetapaper} and the galaxy-galaxy lensing measurements from this work have been analyzed with the same pipeline used in \cite{keypaper}, including the covariance between the two probes and marginalizing over all nuisance parameters like photometric redshift, shear calibration and intrinsic alignments uncertainties. We find the obtained constraints on the galaxy bias from galaxy-galaxy lensing to be in good agreement with those obtained from galaxy clustering.     

\begin{figure}[h]
	\includegraphics[width=\columnwidth]{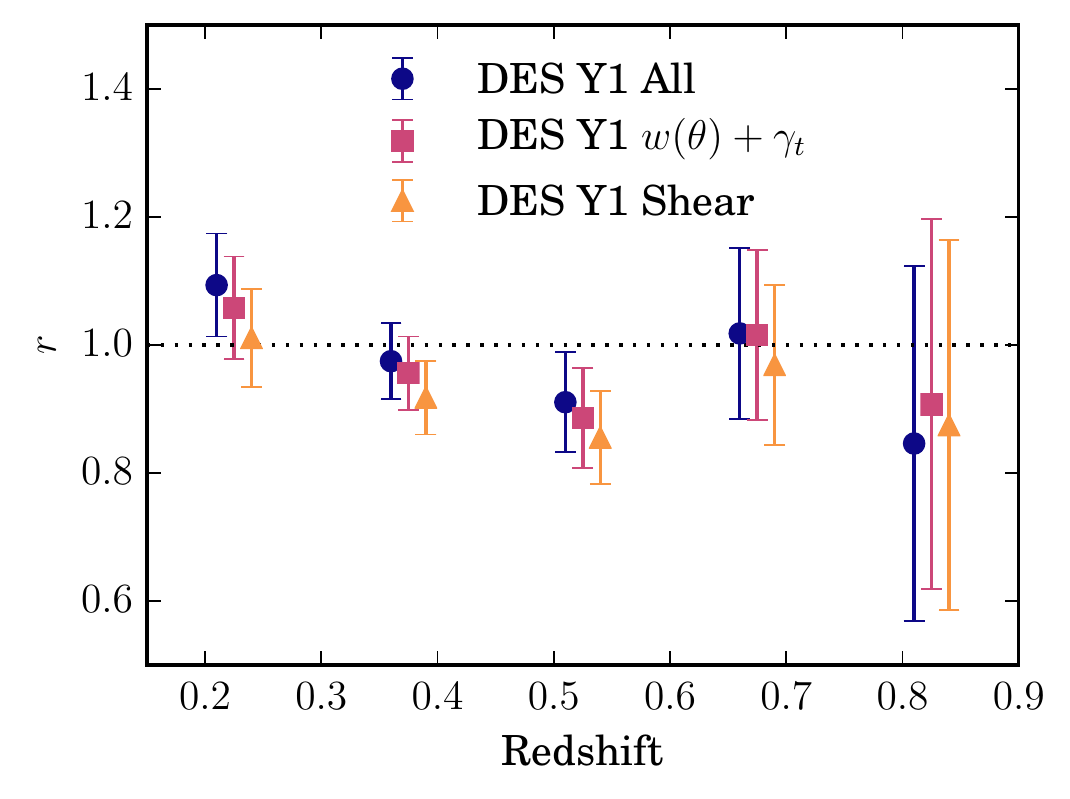}
    \caption{Cross-correlation coefficient $r$ between galaxies and dark matter obtained by comparing the galaxy bias from galaxy clustering ($b_A$) and from galaxy-galaxy lensing only ($b_\times$), fixing all cosmological parameters to three different cosmologies from DES Y1 cosmological results \cite{keypaper}: (i) 3x2 best-fit (All), (ii) $\omega(\theta) + \gamma_t$ best-fit, and (iii) cosmic shear best-fit.}
    \label{fig:r}
\end{figure}

The results in Fig.~\ref{fig:galaxy-bias} can also be interpreted by allowing a non-unity cross-correlation parameter between the galaxy and matter distributions. This parameter is usually expressed in terms of the matter and galaxy power spectra, $P_{\delta \delta}$ and $P_{gg}$ respectively, and the galaxy-matter power spectrum $P_{g\delta}$, as  
\begin{equation} \label{eq: r}
r\,(k, \chi(z)) = \frac{P_{g\delta}(k, \chi(z))}{\sqrt{P_{\delta \delta}(k, \chi(z)) \, P_{gg}(k, \chi(z))}},
\end{equation}
where we have explicitly included its possible scale and redshift dependence. In the context of this model, the galaxy power spectrum remains unchanged with respect to $r=1$, $P_{gg} = b^2 P_{\delta \delta}$, but the galaxy-matter power spectrum changes from $P_{g\delta} = b P_{\delta \delta}$ to $P_{g\delta} = b \, r P_{\delta \delta}$. That introduces an $r$ factor in the galaxy-galaxy lensing expression in Eq.~(\ref{eq:cosmo}), and hence the two estimates of the galaxy bias in Fig.~\ref{fig:galaxy-bias} can be transformed to: 
\begin{equation} 
b = b_A \: ; \: r = b_\times/b_A,
\end{equation} 
and this allows us to place constraints on the $r$ parameter using our measurements. If $r^i$ refers to the cross-correlation parameter in lens bin $i$, the constraints we obtain read: \redd{$r^1 = 1.094 \pm 0.080$, $r^2 = 0.975 \pm 0.059$, $r^3 = 0.911 \pm 0.078$, $r^4 = 1.02 \pm 0.13$, $r^5 = 0.85 \pm 0.28$,} shown also in Fig.~\ref{fig:r}.

In addition, it is important to note that the specified constraints on the galaxy bias and the cross-correlation coefficient are not independent of the assumed cosmology. The values given above are obtained with the 3x2 best-fit cosmological parameters from the DES Y1 main cosmological analysis \cite{keypaper}, which favours the cross-correlation coefficient being consistent with one, since the cosmology is determined assuming the galaxy bias for galaxy clustering and for galaxy-galaxy lensing is the same. This is also true for the 2x2 cosmology, from $\omega(\theta) + \gamma_t$. On the contrary, the cosmological parameters obtained only from the cosmic shear analysis are independent of the galaxy bias and the cross-correlation coefficient and therefore provide a way to test the $r=1$ assumption. In Fig.~\ref{fig:r}, we present the $r$ constraints for each of these three cosmologies, which we find all to be consistent with $r=1$. The $r$ constraints presented in this section provide further justification for assuming $r=1$ in the main DES Y1 cosmological analysis.

In the past, different studies have analyzed the consistency between different estimates of the galaxy bias of a given galaxy population. In the context of DES, a number of different analyses using galaxy clustering in \cite{Crocce2015}, CMB lensing in \cite{Giannantonio2015}, galaxy-galaxy lensing in \cite{Prat2016}, and projected mass maps in \cite{Chang2016} used DES Science Verification (SV) data to obtain constraints on the galaxy bias of the main galaxy population (so-called DES-SV Benchmark sample), finding mild differences in those estimates that were explored as potential differences between clustering and lensing. Outside DES, other studies have also examined potential differences between clustering and lensing. In particular, in \cite{Leauthaud2017} the authors perform a galaxy-galaxy lensing measurement around BOSS CMASS spectroscopic galaxies using data from the CFHTLenS and SDSS Stripe 82 surveys, and find the lensing signal to be lower than that expected from the clustering of lens galaxies and predictions from standard models of the galaxy-halo connection. In this study, as expressed in the $r$ values reported above, and more broadly in the DES Y1 cosmological analysis presented in \cite{keypaper}, we find the clustering and lensing signals to be consistent within our uncertainties, though we note that the \cite{Leauthaud2017} analysis was done on significantly smaller scales.

\section{Conclusions}\label{sec:conclusions}

This paper is part of the Dark Energy Survey Year~1 (DES Y1) effort to obtain cosmological constraints by combining three different probes, namely galaxy clustering, galaxy-galaxy lensing and cosmic shear. The main goal of this work is to present and characterize one of these two-point correlations functions, the galaxy-galaxy lensing measurement. Besides this principal task, we use source tomography to put constraints on the mean of the source redshift distributions using the geometrical shear-ratio test. Finally, we obtain the galaxy bias from this probe and we compare it to the corresponding result from galaxy clustering. 

Our lens sample is composed of redMaGiC galaxies \cite{Rozo2015}, which are photometrically selected luminous red galaxies (LRGs) with high-precision photometric redshifts. This allows us to divide the lens sample into five equally-spaced tomographic bins between 0.15 and 0.9 in redshift. Regarding the source sample, we use two independent shear catalogs, namely \metacal~and \im3shape, which are described in detail in \cite{shearcat}. We split the source galaxies into four tomographic bins between 0.2 and 1.3 in redshift using BPZ, a template-based photometric redshift code. 

In order to characterize the DES Y1 galaxy-galaxy lensing measurements, we test them for an extensive set of potential systematic effects. First, we show that the cross-component of the shear is compatible with zero, which should be the case if the shear is only produced by gravitational lensing. Second, PSF residuals are considered and found to leave no imprint on the tangential shear measurements. Next, we split the source sample into halves of high and low signal-to-noise or size, observing no significant differences between the measurements in each half of the split. Finally, we study the impact of the survey observing conditions, i.e. airmass, seeing, magnitude limit and sky brightness, on the galaxy-galaxy lensing signal, finding no significant dependence. To estimate the significance of these tests we use covariance matrices obtained from the jackknife method, which we validate using a suite of log-normal simulations. Overall, we find no significant evidence of systematics contamination of the galaxy-galaxy lensing signal. Besides serving as crucial input and validation for the DES Y1 cosmological analysis, this set of systematics tests will also be useful for potential future work relying on DES Y1 galaxy-galaxy lensing measurements. 

In addition to the systematics testing, we apply the shear-ratio test to our source tomographic measurements. Given a fixed lens bin, we make use of the geometrical scaling of the tangential shear for different source redshift bins to constrain the mean of the source tomographic redshift distributions, which is one of the dominant sources of uncertainty in the DES Y1 cosmological analysis. For this test, we restrict the scales to those ignored in the cosmological analysis, so that it is independent of the constraints obtained there. Our results are in agreement with other photo-$z$ studies on the same data sample \cite{photoz,xcorr,xcorrtechnique}, thus showing the robustness of the photometric redshifts used in the DES Y1 cosmological analysis. We also find this method to be informative of multiplicative shear biases in the data, hence showing potential as a way of self-calibrating shear biases in future data sets.   

Finally, restricting to the scales used in the cosmological analysis, we use the galaxy-galaxy lensing measurements in this work to obtain galaxy bias constraints on the redMaGiC galaxy sample by fixing all the cosmological parameters but leaving free the nuisance parameters as in \cite{keypaper}. We compare these constraints from the ones obtained using the corresponding galaxy clustering measurements in the same lens sample in \cite{wthetapaper} and using the same cosmological model, finding good agreement between them. This agreement can also be understood as a consistency test of the assumption that the galaxy-matter cross-correlation coefficient $r=1$, made in the cosmology analysis.

\section*{Acknowledgements}

This paper has gone through internal review by the DES collaboration. It has been assigned DES paper id DES-2016-0210 and FermiLab preprint number PUB-17-277-AE.

Funding for the DES Projects has been provided by the U.S. Department of Energy, the U.S. National Science Foundation, the Ministry of Science and Education of Spain,
the Science and Technology Facilities Council of the United Kingdom, the Higher Education Funding Council for England, the National Center for Supercomputing
Applications at the University of Illinois at Urbana-Champaign, the Kavli Institute of Cosmological Physics at the University of Chicago,
the Center for Cosmology and Astro-Particle Physics at the Ohio State University,
the Mitchell Institute for Fundamental Physics and Astronomy at Texas A\&M University, Financiadora de Estudos e Projetos,
Funda{\c c}{\~a}o Carlos Chagas Filho de Amparo {\`a} Pesquisa do Estado do Rio de Janeiro, Conselho Nacional de Desenvolvimento Cient{\'i}fico e Tecnol{\'o}gico and
the Minist{\'e}rio da Ci{\^e}ncia, Tecnologia e Inova{\c c}{\~a}o, the Deutsche Forschungsgemeinschaft and the Collaborating Institutions in the Dark Energy Survey.

The Collaborating Institutions are Argonne National Laboratory, the University of California at Santa Cruz, the University of Cambridge, Centro de Investigaciones Energ{\'e}ticas,
Medioambientales y Tecnol{\'o}gicas-Madrid, the University of Chicago, University College London, the DES-Brazil Consortium, the University of Edinburgh,
the Eidgen{\"o}ssische Technische Hochschule (ETH) Z{\"u}rich,
Fermi National Accelerator Laboratory, the University of Illinois at Urbana-Champaign, the Institut de Ci{\`e}ncies de l'Espai (IEEC/CSIC),
the Institut de F{\'i}sica d'Altes Energies, Lawrence Berkeley National Laboratory, the Ludwig-Maximilians Universit{\"a}t M{\"u}nchen and the associated Excellence Cluster Universe,
the University of Michigan, the National Optical Astronomy Observatory, the University of Nottingham, The Ohio State University, the University of Pennsylvania, the University of Portsmouth,
SLAC National Accelerator Laboratory, Stanford University, the University of Sussex, Texas A\&M University, and the OzDES Membership Consortium.

Based in part on observations at Cerro Tololo Inter-American Observatory, National Optical Astronomy Observatory, which is operated by the Association of
Universities for Research in Astronomy (AURA) under a cooperative agreement with the National Science Foundation.

The DES data management system is supported by the National Science Foundation under Grant Numbers AST-1138766 and AST-1536171.
The DES participants from Spanish institutions are partially supported by MINECO under grants AYA2015-71825, ESP2015-88861, FPA2015-68048, SEV-2012-0234, SEV-2016-0597, and MDM-2015-0509,
some of which include ERDF funds from the European Union. IFAE is partially funded by the CERCA program of the Generalitat de Catalunya.
Research leading to these results has received funding from the European Research
Council under the European Union's Seventh Framework Program (FP7/2007-2013) including ERC grant agreements 240672, 291329, and 306478.
We  acknowledge support from the Australian Research Council Centre of Excellence for All-sky Astrophysics (CAASTRO), through project number CE110001020.

This manuscript has been authored by Fermi Research Alliance, LLC under Contract No. DE-AC02-07CH11359 with the U.S. Department of Energy, Office of Science, Office of High Energy Physics. The United States Government retains and the publisher, by accepting the article for publication, acknowledges that the United States Government retains a non-exclusive, paid-up, irrevocable, world-wide license to publish or reproduce the published form of this manuscript, or allow others to do so, for United States Government purposes.

Support for DG was provided by NASA through Einstein Postdoctoral Fellowship grant
number PF5-160138 awarded by the Chandra X-ray Center, which is
operated by the Smithsonian Astrophysical Observatory for NASA under
contract NAS8-03060. JB acknowledges support from the Swiss National Science Foundation.

This research used computing resources at SLAC
National Accelerator Laboratory, and at the National Energy
Research Scientific Computing Center, a DOE Office of Science User
Facility supported by the Office of Science of the U.S. Department of
Energy under Contract No. DE-AC02-05CH11231.




\bibliographystyle{apsrev4-1}  
\bibliography{library}


\appendix
\section{$\Delta \Sigma$ and $\gamma_t$}
\label{sec:deltasigma}

When we measure the mean tangential alignment of background galaxies around lenses, we need to make a choice as to how we weight each of the lens-source pairs. In this appendix, we discuss the implications of using either a uniform weight for all source-lens pairs in a given combination of source and lens redshift bins, or a weight that takes into account the photometric redshift estimate of the source to yield a minimum variance estimate of the surface mass density contrast of the lens.

In the first case, 
and without a shape noise weighting of sources, 
our measurement $\gamma_t$ is simply the arithmetic mean of the tangential components of ellipticities of sources $i$:
\begin{equation}
\gamma_t=N^{-1}\sum_{i=1}^N e_{t,i} \; .
\label{eqn:gmean}
\end{equation}

In the second case, we weight each lens-source pair by a weight $w_{\mathrm{e},i}$,
\begin{equation}
\gamma_t=\frac{\sum_{i=1}^N w_{\mathrm{e},i} e_{t,i}}{\sum_{i=1}^N w_{\mathrm{e},i}} \; .
\label{eqn:gwmean}
\end{equation}
For optimal signal-to-noise ratio and uniform shape noise of our sample of source galaxies, $w_{\mathrm{e},i}$ should be chosen to be proportional to the amplitude of the signal in each lens-source pair, i.e.
\begin{equation}\label{eqn:optweight}
w_{\mathrm{e},i}\propto\frac{D_{\rm l} D_{\rm ls}}{D_{\rm s}} \; .
\end{equation}

We note that, for a given cosmology, the mean shears of both Eq.~(\ref{eqn:gmean}) and Eq.~(\ref{eqn:gwmean}) can be converted to an estimate of surface mass density $\Delta\Sigma$, by multiplying with the (weighted) estimate of $\Sigma_{\rm crit}^{-1}$, as in Eq.~(\ref{eq:gammat_delta_sigma}). In the case of Eq.~(\ref{eqn:gwmean})
with the weights equal to the expectation value of Eq.~(\ref{eqn:optweight}), this is identical to the common $\Delta\Sigma$ estimator of \cite{Sheldon2004}.

The unweighted mean of Eq.~(\ref{eqn:gmean}) has the considerable advantage that nuisance parameters describing the systematic uncertainty of shear and redshift estimates of the source redshift bins are identical to the ones determined for a cosmic shear analysis using the same samples \cite{shearcorr, photoz, shearcat}. This is of particular importance when joining cosmic shear and galaxy-galaxy lensing measurements into one combined probe \cite{keypaper}. The question at hand therefore is whether the increase in signal-to-noise ratio (S/N) due to the optimal weighting of Eq.~(\ref{eqn:gwmean}) would warrant the added complication.

We make a simple estimate of the loss in S/N incurred by uniform weighting of sources. To this end, we simulate a source sample with overall Gaussian distribution of true redshifts $z_t$ with a mean $\langle z_t\rangle=0.6$ and width $\sigma_{t}=0.3$. We split sources into redshift bins of width $\Delta z_p=0.25$ by a point estimate $z_p$ of their redshift. For a given source redshift bin centered on $z_m$, we emulate the latter by adding a Gaussian scatter of $\sigma_p =-0.1\,(1+z)+0.12\,(1+z)^2$ to $z_t$, which is a realistic scatter for DES-Y1 photo-$z$'s \cite{photoz}.

\begin{figure}
	\includegraphics[width=0.48\textwidth]{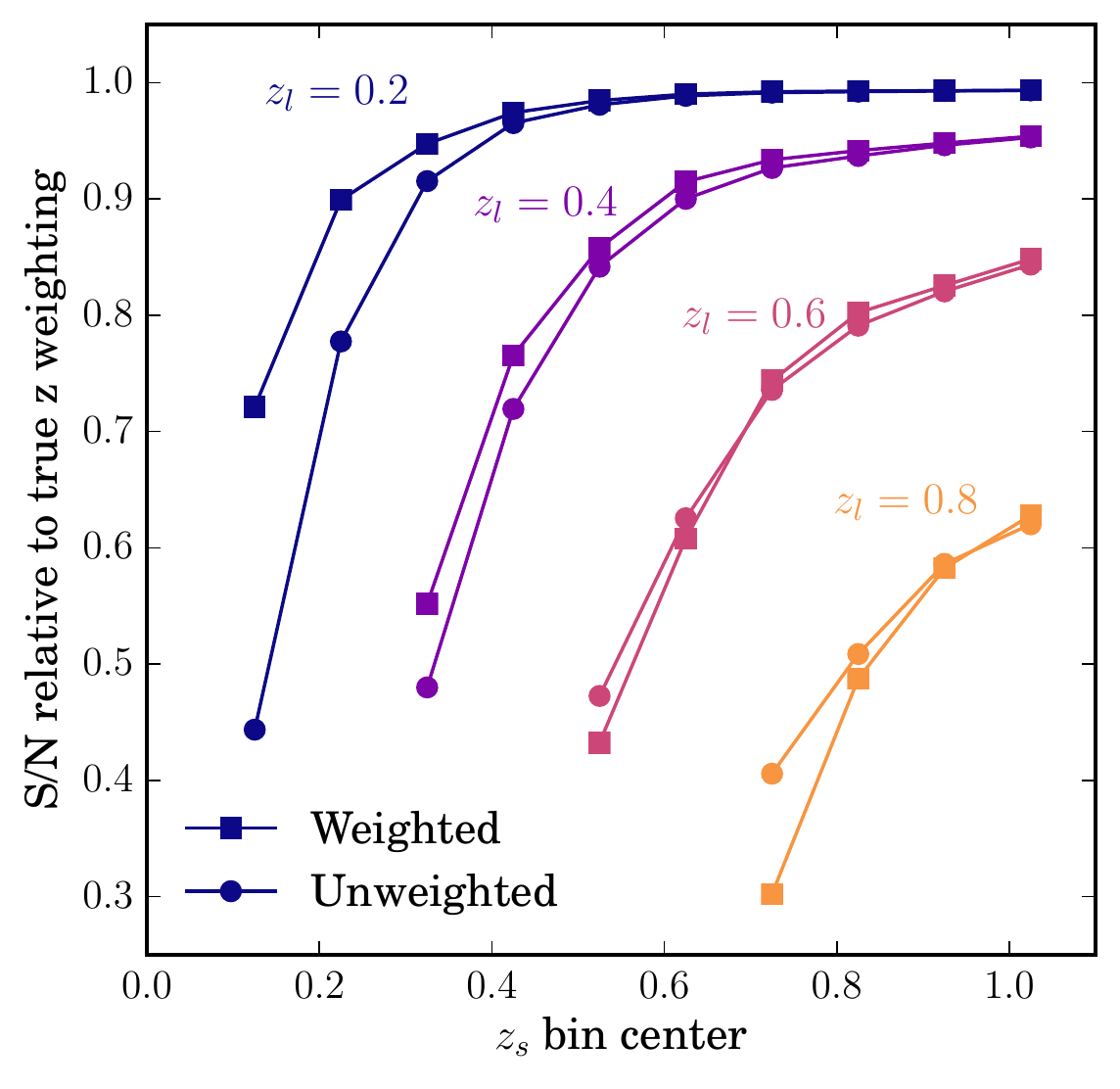}
    \caption{Relative signal-to-noise ratio of lensing signal recovered when weighting sources uniformly (commonly called $\gamma_{\rm t}$, circles) and with "$\Delta\Sigma$ weighting" according to a DES-like photometric redshift point estimate (squares) with $\sigma_p =-0.1\,(1+z)+0.12\,(1+z)^2$ scatter around the true redshift \cite{photoz}. The point estimate is used to select source bins of width $\Delta z=0.25$.}
    \label{fig:ds_sn}
\end{figure}

Figure~\ref{fig:ds_sn} compares the recovered S/N of the galaxy-galaxy lensing signal to that of weighting each source by the optimal weight using its true redshift for two cases: (1) uniform weighting of all sources in a redshift bin (circles) and (2) weighting each source by Eq.~(\ref{eqn:optweight}) evaluated at the source redshift point estimate (squares). Except in the case of source redshift bins overlapping the lens redshift, uniform weighting does not considerably lower the S/N of the measured galaxy-galaxy lensing signal. The photo-$z$ resolution results in a bigger gain when using optimal weighting compared to uniform weighting. For instance, for $z_l=0.4$ and $z_s=0.425$, the gain of using photo-$z$ optimal weighting is 6.4\% for the fiducial photo-$z$ scatter while it goes up to $25\%$ if we improve the resolution by a factor of two. In a case with less overlap between the lens and source redshift distributions the improvement is reduced, as expected. For example, for $z_l=0.4$ and $z_s=0.625$, the gain of using photo-$z$ optimal weighting is 1.6\% for the fiducial photo-$z$ scatter while it is 2.1\% for a photo-$z$ resolution that is twice as good. Therefore, we conclude that, even though optimal weighting can be important, for the photo-$z$ precision and the source binning used in this work, photo-$z$-dependent weighting of sources does not significantly improve the constraining power, and decide to use uniformly weighted tangential shears in this analysis.

\section{Effect of random point subtraction in the tangential shear measurement}\label{sec:appendix_rp}

\begin{figure*}
	\includegraphics[width=1\textwidth]{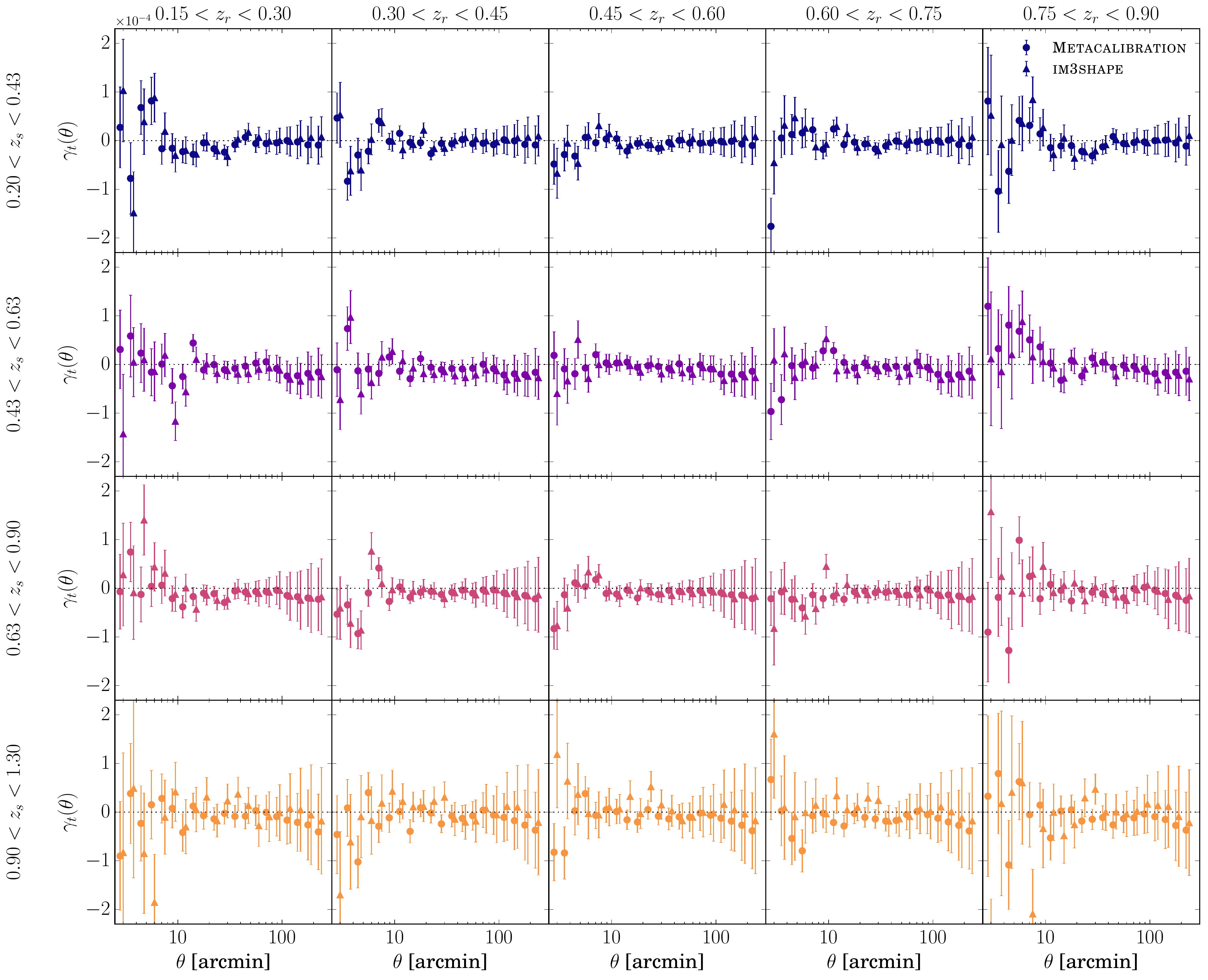}
    \caption{Tangential shear around random points for \textsc{Metacalibration} and \textsc{im3shape}.}
    \label{fig:randoms}
\end{figure*}

\begin{figure*}
	\includegraphics[width=0.6\textwidth]{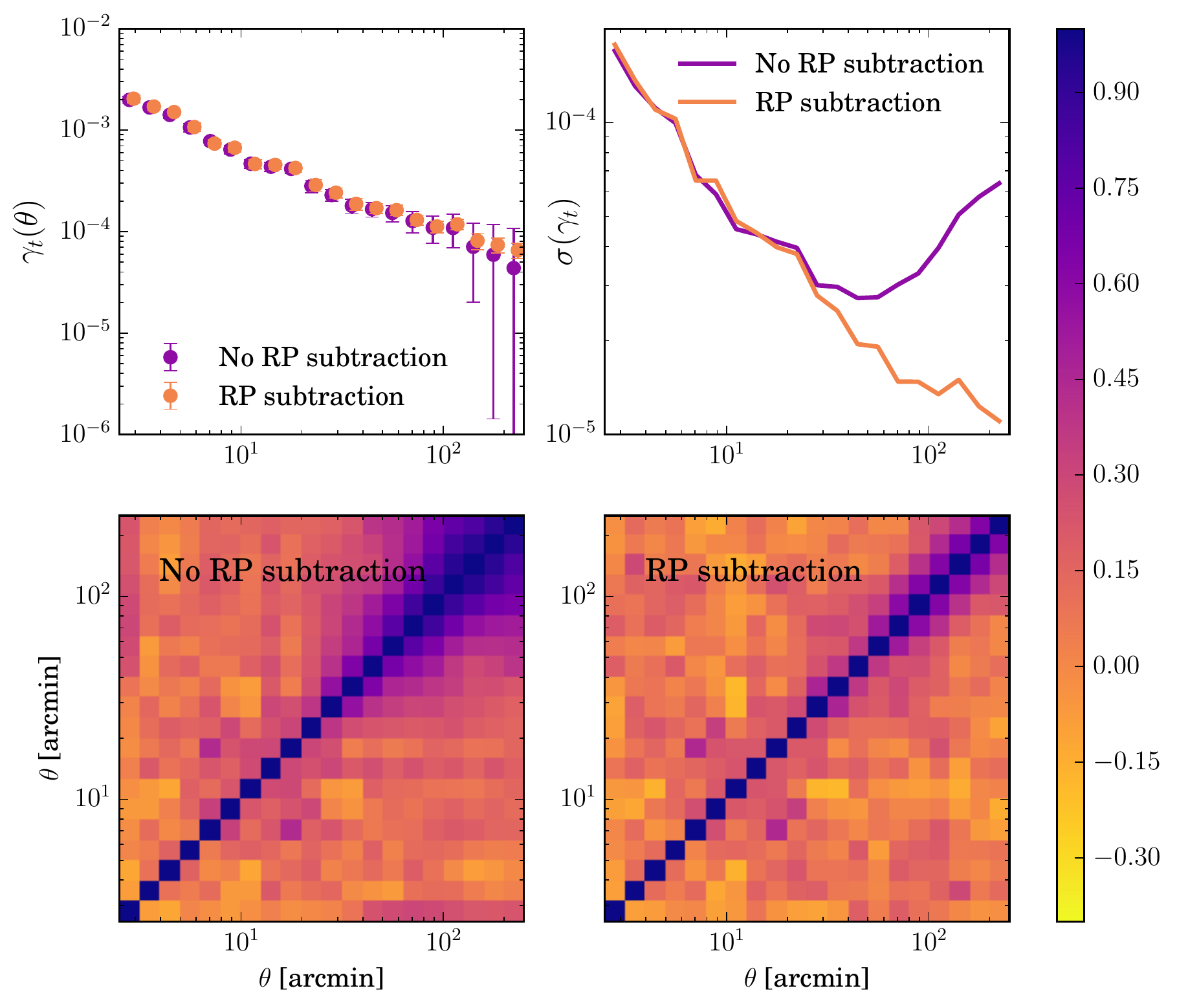}
    \caption{We show the impact the random point subtraction has on the tangential shear measurement and its corresponding jackknife covariance matrix for an example redshift bin ($0.3<z_l<0.45$ and $0.63<z_s<0.90$ for \metacal).}
    \label{fig:cov}
\end{figure*}

Our estimator of galaxy-galaxy lensing in Eq.~(\ref{eq: random points subtraction}) includes subtracting the measurement around random points that trace the same survey geometry. This measurement, using a set of random points with 10 times as many points as lens galaxies, is shown in Fig.~\ref{fig:randoms}. Even though this is a correction included in the measurement, it is nonetheless useful to confirm that it is small at all scales used in the analysis. The measurement tests the importance of systematic shear which is especially problematic at the survey boundary, and allows us to compare the magnitude of the systematic shear with the magnitude of the signal around actual lens galaxies. We find the tangential shear around random points to be a small correction, consistent with the null hypothesis, as it is seen in the top left panel of Fig.~\ref{fig:cov}.

Even though the random point subtraction is a mild correction to the signal, it has an important effect on the covariance matrix. Subtracting the measurement around random points removes a term in the covariance due to performing the measurement using the over-density field instead of the density field, as it was studied in detail in \cite{Singh2016}. As seen in Fig.~\ref{fig:cov}, we observe this effect on scales larger than 20 arcmin., where the covariance is no longer dominated by shape noise. When subtracting the measurement around random points, we detect both a significant decrease on the uncertainty of the tangential shear (top right panel) and a reduction of the correlation between angular bins (lower panels).

Finally, another argument that strongly favours applying the random points subtraction is the following. In Sec.~\ref{subsec:cov_matrix} we validated the jackknife method using log-normal simulations, showing that the uncertainties on the tangential shear are compatible when using the jackknife method and when using the true variance from 1200 independent FLASK simulations (Fig.~\ref{fig:sigma_gammat_sims_comparison}). We have performed this comparison both with and without the random point subtraction, finding that there is only agreement between the different methods when the tangential shear around random points is removed from the signal.

\section{\metacal~responses scale dependence}\label{sec:appendix_responses}

As explained in Sec.~\ref{sec:metacal_responses}, when applying the \metacal~responses we approximate them as being scale independent. In this appendix we test the validity of this approximation by measuring the scale dependence of the responses for all the tomographic lens-source bin combinations. 

\begin{figure*}
	\includegraphics[width=1\textwidth]{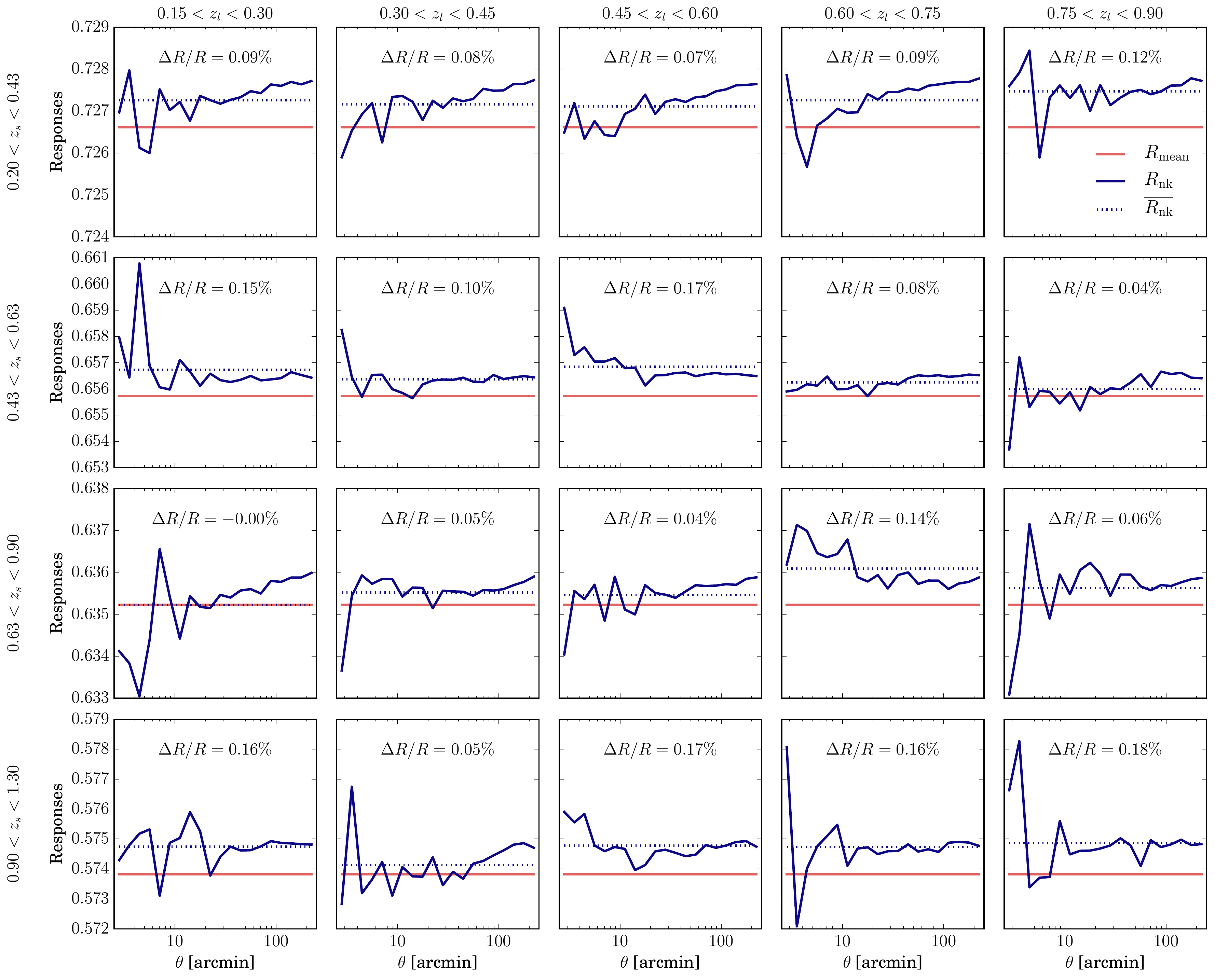}
    \caption{\metacal~responses scale dependence and mean values. We compare the responses averaged in 20 log-spaced angular bins between 2.5 and 250 arcmin in each lens-source redshift bin combination ($R_{\mathrm{nk}}$) to the average of the responses in each source redshift bin ($R_\mathrm{mean}$). The maximum difference between them is at the $0.2\%$ level.}
    \label{fig:responses}
\end{figure*}

In Fig.~\ref{fig:responses} we display the \metacal~responses for all the lens-source redshift bins combinations averaged in 20 log-spaced angular bins using the \texttt{NK} \texttt{TreeCorr} correlation function. Comparing to the mean of the responses over the ensemble in each source redshift bin, we find the variation with $\theta$ to be very small compared to the size of our measurement uncertainties and thus decide to use a constant value for simplicity. Future analyses using \metacal~on larger data samples with smaller uncertainties may need to include the scale-dependent responses in their measurements.

\end{document}